\documentclass[12pt]{article}
\usepackage{graphicx}
\usepackage{fancybox}
\usepackage{fancyhdr}
\usepackage{amsmath}
\usepackage{amssymb}
\usepackage{epsfig}
\usepackage{verbatim}
\usepackage{color}
\usepackage{fancybox}
\usepackage{euscript}
\usepackage{cite}
\usepackage{epic}
\newcommand{\mathswitchr}[1]{\relax\ifmmode{\mathrm{#1}}\else$\mathrm{#1}$\fi}
\newcommand{\Ei}{\mathop{\mathrm{Ei}}\nolimits}

\newcommand{\FYFS}{F_{\mathrm YFS}}
\newcommand{\rQCED}{\mathrm {QCED}}
\begin{document}
\begin{titlepage}
\begin{flushleft}BU-HEPP-09-07\\
Oct., 2009
\end{flushleft}
\begin{center}
{\Large New Approach to Parton Shower MC's for Precision QCD Theory: HERWIRI1.0(31)}
\vspace{2mm}

{\bf S. Joseph$^a$, S. Majhi$^b$, B.F.L. Ward$^a$, S. A. Yost$^c$}\\
\vspace{2mm}
{\em $^a$Department of Physics, Baylor University, Waco, TX 76798, USA}\\
%\author{B.F.L. Ward}
%\affiliation{Department of Physics, Baylor University, Waco, TX 76798, USA}
%\author{S. Majhi}
{\em $^b$Theory Division, Saha Institute of Nuclear Physics, Kolkata 700 064, India}\\
%\author{B.F.L. Ward}
%\affiliation{Department of Physics, Baylor University, Waco, TX 76798, USA}
%\author{S. A. Yost}
{\em $^c$Department of Physics, The Citadel, Charleston, SC 29409, USA}\\
%
%\vspace{3mm}
\end{center}
\vspace{3mm}
\begin{center}
{\bf Abstract}
\end{center}
By implementing the new IR-improved 
Dokshitzer-Gribov-Lipatov-Altarelli-Parisi-Callan-Symanzik (DGLAP-CS) kernels recently developed by one of us in
the HERWIG6.5 environment we generate a new MC, HERWIRI1.0(31),
for hadron-hadron scattering at high energies.
We use MC data to illustrate 
the comparison between the parton shower generated by the 
standard DGLAP-CS kernels and that generated by the new IR-improved 
DGLAP-CS kernels. The interface to MC@NLO, MC@NLO/HERWIRI, is illustrated.
Comparisons with FNAL data
and some discussion of possible 
implications for LHC phenomenology are also presented.
%This document serves as a template for the proceedings of the DPF-2009 conference.  
%Authors should prepare their papers using a copy and follow the guidelines described here.
%Please do not modify the page layout or styles. 
%\end{abstract}
\vspace{10mm}
\renewcommand{\baselinestretch}{0.1}
\footnoterule
%\maketitle
\end{titlepage}
\def\Kmax{K_{\rm max}}\def\ieps{{i\epsilon}}\def\rQCD{{\rm QCD}}
\renewcommand{\theequation}{\arabic{equation}}
\font\fortssbx=cmssbx10 scaled \magstep2
\renewcommand\thepage{}
%\vfill\eject
\parskip.1truein\parindent=20pt\pagenumbering{arabic}\par
%\newpage
\section{\label{intro} Introduction}
In the era of the LHC, we must deal with requirements of
precision QCD, which entails  
predictions for QCD processes at the total precision~\cite{jadach1} tag
of $1\%$ or better, where by total precision of a theoretical 
prediction we mean the technical and physical 
precisions combined in quadrature 
or otherwise as appropriate. We accordingly need resummed 
${\cal O}(\alpha_s^2L^n),{\cal O}(\alpha_s\alpha L^{n'}),
{\cal O}(\alpha^2 L^{n''})$ corrections for $n=0,1,2,$ $n'=0,1,2,$ $n''=1,2$, 
in the presence of parton showers, on an event-by-event basis, 
without double counting and with exact phase space. Essential 
large QED and EW effects~\cite{qedeffects,radcor-ew,ditt-lp09} are handled
by the simultaneous
resummation of large QED and QCD infrared(IR) 
effects, QED$\otimes$QCD resummation
~\cite{qced} in the presence of parton showers, to be realized on an 
event-by-event basis by Monte Carlo (MC) event generator
methods. Indeed, we know from
Refs.~\cite{radcor-ew,ditt-lp09} that 
no precision prediction for a hard LHC process
at the 1\% level can be complete without taking the large 
EW corrections into account. \par\indent

In what follows, we present the first step in realizing our 
new MC event generator approach to precision LHC physics
with amplitude-based QED$\otimes$QCD resummation by introducing the attendant
new parton shower MC for QCD that follows from our approach. We recall
that in Refs.~\cite{jad-ward} our resummed QED MC methods, based on the 
theory in Ref.~\cite{yfs},
are already well developed and checked in LEP1 and LEP2
precision physics applications. This means that what we do here will set 
the stage for the complete implementation, via MC methods,
the QED$\otimes$QCD resummed theory in which all IR singularities are
canceled 
to all orders in $\alpha_s$ and $\alpha$.
As we will show directly, this new parton shower MC, which is developed 
in the HERWIG6.5~\cite{herwig} environment and which we have
called HERWIRI1.0(31)~\cite{irdglap3-plb}\footnote{We stress that we are completely replacing all the parton shower
kernels in HERWIG6.5 that generate real QCD radiation and all the Sudakov form factors in HERWIG6.5 that realize the attendant virtual corrections with the new forms that follow from the results in Sections 2 and 3,
together with the auxiliary functions required for these new forms, so that the parton shower physics, distributions and MC behavior are all fundamentally different from what is in HERWIG6.5. The 
attendant implementation has been carried out in agreement with and in close collaboration with B. Webber and M. Seymour, principal authors of the HERWIG6.5, who have also instructed us on the proper naming and references for the resulting program as we further highlight in the discussion below.}, already shows improvement in
comparison with the FNAL soft $p_T$ data on single $Z$ production
as we quantify below.
On the theoretical side, while the explicit IR cut-offs in the HERWIG6.5
environment will not be removed here, our new shower MC 
only involves integrable
distributions for its real emission
so that in principle these cut-offs could be removed. 
We discuss this point further below as well.\par\indent

Our discussion here proceeds as follows. We first review
our approach to resummation and its relationship to those in Refs.~\cite{cattrent,scet}. Section 3 contains a presentation of the attendant new IR-improved DGLAP-CS~\cite{dglap,cs} theory~\cite{irdglap1,irdglap2}.
Section 4 features the implementation of the new IR-improved kernels in the framework of HERWIG6.5~\cite{herwig} to arrive at the new, IR-improved parton shower MC
HERWIRI1.0(31). We illustrate the effects of the IR-improvement first with the 
generic 2$\rightarrow$2 processes at LHC energies and then  with the specific
single $Z$ production process at LHC energies. We compare with recent 
data from FNAL to make
direct contact with observation. 
Section~\ref{concl} summarizes our discussion.
\par\indent %\vskip0.2cm

For reference purposes and to put the discussion in the proper perspective
with regard to what has already been achieved in the relevant literature, 
we note that the authors
%analyses 
in Refs.~\cite{scott1,scott2}
%. The authors in the latter references
have argued that
the current state-of-the-art theoretical precision tag on single $Z$
production at the LHC is 
$(4.1\pm0.3)\%=(1.51\pm 0.75)\%(QCD)\oplus 3.79(PDF)\oplus 0.38\pm 0.26(EW)\%$ 
and that the analogous estimate for single $W$ production is $\sim 5.7$\%. 
We continue to emphasize that these estimates show how
much more work is still needed to achieve the desired 1.0\% total precision tag
on these two processes, for example.\par\indent

%,
%where the results of Refs.~\cite{cteq,mrst,mcnlo,fewz,resbos,horace,photos} have been
%used.
%\footnote{Recently, the 
%analogous estimate for single W production has been given ~\cite{scott2} as $\sim 5.7$\%.}\par%\vskip0.2cm
%%%
%%%Start Here
\section{QED$\otimes$QCD Resummation}
We follow here the discussion in Refs.~\cite{qced,irdglap1,irdglap2}, wherein
we have derived the following expression for the 
hard cross sections in the SM $SU_{2L}\times U_1\times SU_3^c$ EW-QCD theory
\begin{eqnarray}
d\hat\sigma_{\rm exp} &=& e^{\rm SUM_{IR}(QCED)}
   \sum_{{n,m}=0}^\infty\frac{1}{n!m!}\int
\frac{d^3p_2}{p_2^{0}}\frac{d^3q_2}{q_2^{0}}
\prod_{j_1=1}^n\frac{d^3k_{j_1}}{k_{j_1}} 
\prod_{j_2=1}^m\frac{d^3{k'}_{j_2}}{{k'}_{j_2}}
\nonumber\\
& & \kern-2cm  \times \int\frac{d^4y}{(2\pi)^4}
e^{iy\cdot(p_1+q_1-p_2-q_2-\sum k_{j_1}-\sum {k'}_{j_2})+ D_\rQCED}
 \tilde{\bar\beta}_{n,m}(k_1,\ldots,k_n;k'_1,\ldots,k'_m),
\label{subp15b}
\end{eqnarray}
where the new YFS-style~\cite{yfs} residuals
$\tilde{\bar\beta}_{n,m}(k_1,\ldots,k_n;k'_1,\ldots,k'_m)$ have $n$ hard gluons and $m$ hard photons and we show the final state with two hard final
partons with momenta $p_2,\; q_2$ specified for a generic 2f final state for
definiteness. 
The infrared functions ${\rm SUM_{IR}(QCED)},\; D_\rQCED$
are defined in Refs.~\cite{qced,irdglap1,irdglap2}. Eq. (\ref{subp15b}) 
is an exact
implementation of amplitude-based simultaneous resummation 
of QED and QCD large IR effects valid to all orders 
in $\alpha$ and in $\alpha_s$. When restricted to its QED aspect, it is the basis of the well established YFS MC approach~\cite{jad-ward} to precision multiple photon radiative effects that is well tested already in LEP1 and LEP2 precision physics applications. Thus what we present in this paper, the first 
realization
of the new parton shower MC for QCD that follows from the QCD aspect of 
(\ref{subp15b}), opens the way to the full MC implementation of all aspects
of our QED$\otimes$QCD resummatiom theory approach to precision LHC physics predictions.\par\indent

The approach to QCD resummation contained in (\ref{subp15b}) is fully consistent with that of
Refs.~\cite{cattrent,scet} as follows. First, Ref.~\cite{geor1} has shown that the latter two approaches are equivalent. We show in Refs.~\cite{irdglap1,irdglap2}
that our approach is consistent with that of Refs.~\cite{cattrent}
by exhibiting the transformation prescription from the resummation formula
for the theory in Refs.~\cite{cattrent} for the generic $2\rightarrow n$ parton process as given in Ref.~\cite{madg} to our theory as given for QCD by restricting Eq.\ (\ref{subp15b}) to its QCD component, where a key point is to use the color-spin density matrix formulation of our residuals to capture the respective full quantum mechanical color-spin correlations in the results in Ref.~\cite{madg}. For completeness, we let us recapitulate the essence of
the attendant discussion here, as the arguments are not generally well-known.
More precisely, to illustrate the relationship between our approach and that in 
Refs.~\cite{cattrent}, we use as a vehicle Ref.~\cite{madg}, 
which treats the $2\rightarrow n$ parton process in the resummation
theory of Refs~\cite{cattrent}, working in the IR and collinear regime
to exact 2-loop order. The authors in Ref.~\cite{madg} have arrived at the
following representation for the amplitude for
a general 2 $\rightarrow n$ parton process [f] at hard scale Q,
$f_1(p_1,r_1)+f_2(p_2,r_2)\rightarrow f_3(p_3,r_3)+f_4(p_4,r_4)+\cdots+f_{n+2}(p_{n+2},r_{n+2})$, where the $p_i,r_i$ label 4-momenta and color indices respectively, with all parton masses set to zero 
( so in our approach, we should have in mind that
the masses of the quarks (see the discussion below) 
and the IR regulator mass of the gluon 
would all be taken to zero or, we could, as it is done Ref.~\cite{madg}, 
just set all masses to zero at the outset and use dimensional 
regularization to define both collinear and IR singular integrals)
\begin{equation}
\begin{split}
{\cal M}^{[f]}_{\{r_i\}}&=\sum^{C}_{L}{\cal M}^{[f]}_L(c_L)_{\{r_i\}}\\
&= J^{[f]}\sum^{C}_{L}S_{LI}H^{[f]}_I(c_L)_{\{r_i\}},
\end{split}
\label{madg1}
\end{equation}
where repeated indices are summed, and the functions $J^{[f]},S_{LI}$, and $H^{[f]}_I $ are respectively the jet function, the soft function which describes
the exchange of soft gluons between the external lines, and the hard coefficient function. The latter functions' infrared and collinear 
poles have been calculated to 2-loop order in Refs.~\cite{madg}. How do these
results relate to eq.(\ref{subp15b})?\par
To make contact between eqs.(\ref{subp15b},\ref{madg1}), identify in the
specific application
$\bar{Q}'Q\rightarrow \bar{Q}'''Q''+m(G)$ in (\ref{subp15b}) $f_1=Q, f_2=\bar{Q}',
f_3=Q'', f_4=\bar{Q}''', \{f_5,\cdots,f_{n+2}\}=\{G_1,\cdots,G_m\}$,in (\ref{madg1}), where we use the obvious notation for the gluons here. This means that
$n=m+2$. Then, to use eq.(\ref{madg1}) in eq.(\ref{subp15b}), 
one simply has to observe the following:\begin{description}
\item{I.} By its definition in eq.(2.23) of Ref.~\cite{madg}, the anomalous dimension
of the matrix $S_{LI}$ does not contain any of the diagonal effects described by our infrared functions ${\rm SUM_{IR}(QCD)}$ and $D_{\rm QCD}$, where
\[ {\rm SUM_{IR}(QCD)}=2\alpha_s \Re B_{QCD}+2\alpha_s\tilde B_{QCD}(\Kmax),\]
\[ 2\alpha_s\tilde B_{QCD}(\Kmax)=\int{d^3k\over k^0}\tilde S_\rQCD(k)
\theta(\Kmax-k),\]
 \begin{equation} D_\rQCD=\int{d^3k\over k}\tilde S_\rQCD(k)
\left[e^{-iy\cdot k}-\theta(\Kmax-k)\right],\label{subp11a}\end{equation},
where the real IR emission function 
$\tilde S_{\rm QCD}(k)$ and the virtual IR function $\Re B_{QCD}$
are defined eqs.(77,73) in Ref.~\cite{irdglap1}. Note that (\ref{subp15b})
is independent of $K_{max}$.
\item{II.} By its definition in eqs.(2.5) and (2.7) of Ref.~\cite{madg}, the jet function $J^{[f]}$ contains the exponential of the virtual infrared function $\alpha_s\Re{B}_{QCD}$, so that we have to take care that we do not double count when we
use (\ref{madg1}) in (\ref{subp15b}) and in the equations in 
Refs.~\cite{qced,irdglap1,irdglap2} that lead thereto.\end{description}
When we observe these two latter points, we get the following realization
of our approach using the results in Ref.~\cite{madg}:
In our result in eq.(75) in Ref.~\cite{irdglap1} for the contribution
to (\ref{subp15b}) of $m$-hard gluons for the process under study here, 
\begin{eqnarray}
  d\hat\sigma^m = {e^{2\alpha_s\Re B_{QCD}}\over {m !}}\int\prod_{j=1}^m
{d^3k_j\over (k_j^2+\lambda^2)^{1/2}}\delta(p_1+q_1-p_2-q_2-\sum_{i=1}^mk_i)
\nonumber\\       
\bar\rho^{(m)}(p_1,q_1,p_2,q_2,k_1,\cdots,k_m)
{d^3p_2d^3q_2\over p^0_2 q^0_2},
\label{diff1}
\end{eqnarray}
we can identify the residual $\bar\rho^{(m)}$ as follows:{\small
\begin{equation}
\begin{split}\bar\rho^{(m)}(p_1,q_1,p_2,q_2,k_1,\cdots,k_m)
&=\overline\sum_{colors,spin}|{\cal M}^{[f]}_{\{r_i\}}|^2\\
&\kern-2cm\equiv \sum_{spins,\{r_i\},\{r'_i\}}\mathfrak{h}^{cs}_{\{r_i\}\{r'_i\}}|\bar{J}^{[f]}|^2\sum^{C}_{L=1}\sum^{C}_{L'=1}S^{[f]}_{LI}H^{[f]}_I(c_L)_{\{r_i\}}\left(S^{[f]}_{L'I'}H^{[f]}_{I'}(c_{L'})_{\{r'_i\}}\right)^\dagger,
\end{split}
\label{madg2}
\end{equation}}
where here we defined $\bar{J}^{[f]}=e^{-\alpha_s\Re{B}_{QCD}}J^{[f]}$, 
and we introduced the color-spin density matrix for the initial state, $\mathfrak{h}^{cs}$, so that
$\mathfrak{h}^{cs}_{\{r_i\}\{r'_i\}}=\mathfrak{h}^{cs}_{\{r_1,r_2\}\{r'_1,r'_2\}}$, suppressing the spin indices, i.e., $\mathfrak{h}^{cs}$ only depends on the initial state colors and has the obvious normalization implied by (\ref{diff1}). Proceeding then according to
the steps in Ref.~\cite{irdglap1} leading from (\ref{diff1}) to 
(\ref{subp15b}) restricted to QCD, we get the corresponding implementation of the results in Ref.~\cite{madg} in our approach, without
any double counting of effects.\par 
% -- see Refs.~\cite{irdglap1,irdglap2} for details.

\section{IR-Improved DGLAP-CS Theory}
We show in Refs.~\cite{irdglap1,irdglap2} that the result Eq.\ (\ref{subp15b})
restricted to QCD allows us to improve in the IR regime 
%\footnote{This 
%should be distinguished from the also important
%resummation in parton density evolution for the ``$z\rightarrow 0$'' regime,
%where Regge asymptotics obtain -- see for example Ref.~\cite{ermlv,guido}. This
%improvement must also be taken into account for precision LHC predictions.} 
the kernels in DGLAP-CS~\cite{dglap,cs}
theory as follows, using a standard notation:
\begin{align}
P^{exp}_{qq}(z)&= C_F \FYFS(\gamma_q)e^{\frac{1}{2}\delta_q}\left[\frac{1+z^2}{1-z}(1-z)^{\gamma_q} -f_q(\gamma_q)\delta(1-z)\right],\nonumber\\
P^{exp}_{Gq}(z)&= C_F \FYFS(\gamma_q)e^{\frac{1}{2}\delta_q}\frac{1+(1-z)^2}{z} z^{\gamma_q},\nonumber\\
P^{exp}_{GG}(z)&= 2C_G \FYFS(\gamma_G)e^{\frac{1}{2}\delta_G}\{ \frac{1-z}{z}z^{\gamma_G}+\frac{z}{1-z}(1-z)^{\gamma_G}\nonumber\\
&\qquad +\frac{1}{2}(z^{1+\gamma_G}(1-z)+z(1-z)^{1+\gamma_G}) - f_G(\gamma_G) \delta(1-z)\},\nonumber\\
P^{exp}_{qG}(z)&= \FYFS(\gamma_G)e^{\frac{1}{2}\delta_G}\frac{1}{2}\{ z^2(1-z)^{\gamma_G}+(1-z)^2z^{\gamma_G}\},
%P_{qG}(z)&=\frac{1}{2}(z^2+(1-z)^2).
\label{dglap19}
\end{align}
where the superscript ``exp'' indicates that the kernel has been resummed as
predicted by Eq.\ (\ref{subp15b}) when it is restricted to QCD alone and where
\begin{align}
\gamma_q &= C_F\frac{\alpha_s}{\pi}t=\frac{4C_F}{\beta_0}, \qquad \qquad
\delta_q =\frac{\gamma_q}{2}+\frac{\alpha_sC_F}{\pi}(\frac{\pi^2}{3}-\frac{1}{2}),\nonumber\\
f_q(\gamma_q)&=\frac{2}{\gamma_q}-\frac{2}{\gamma_q+1}+\frac{1}{\gamma_q+2},\nonumber\\
\gamma_G &= C_G\frac{\alpha_s}{\pi}t=\frac{4C_G}{\beta_0}, \qquad \qquad
\delta_G =\frac{\gamma_G}{2}+\frac{\alpha_sC_G}{\pi}(\frac{\pi^2}{3}-\frac{1}{2}),\nonumber\\
f_G(\gamma_G)&=\frac{n_f}{6C_G \FYFS(\gamma_G)}{e^{-\frac{1}{2}\delta_G}}+
\frac{2}{\gamma_G(1+\gamma_G)(2+\gamma_G)}+\frac{1}{(1+\gamma_G)(2+\gamma_G)}\\
%+\frac{1}{12}\}.
&\qquad +\frac{1}{2(3+\gamma_G)(4+\gamma_G)}+\frac{1}{(2+\gamma_G)(3+\gamma_G)(4+\gamma_G)},\nonumber\\
\FYFS(\gamma)&=\frac{e^{-C\gamma}}{\Gamma(1+\gamma)}, \qquad \qquad \qquad C=0.57721566... ,
\end{align}
where $\Gamma(w)$ is Euler's gamma function and $C$ is Euler's constant.
We use a one-loop formula for $\alpha_s(Q)$, so that
\[\beta_0=11-\frac{2}{3}n_f,\] where $n_f$ is the number of
active quark flavors and $C_F=4/3$ and $C_G=3$ are the 
respective quadratic Casimir invariants 
for the quark and gluon color representations.\par
\begin{figure}
\begin{center}
\setlength{\unitlength}{1mm}
\begin{picture}(160,80)
%%\put(0,0){\framebox( 65,60){ }}
\put(-2.4, -10){\makebox(0,0)[lb]{
%\epsfig{file=epi02-fg1g.eps,width=80mm,height=30mm}
%}}
\epsfig{file=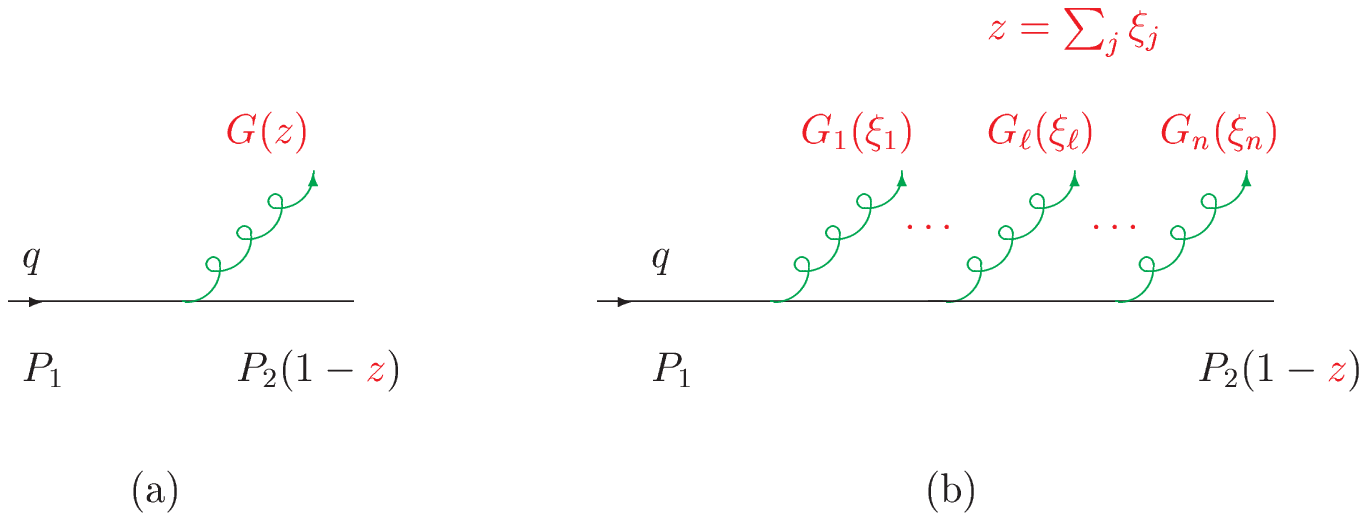,width=140mm}
}}
\end{picture}
\end{center}
%\label{fig1-a}
\caption{In (a), we show the usual process $q\rightarrow q(1-z)+G(z)$;
in (b), we show its multiple gluon improvement $q\rightarrow q(1-z)+G_1(\xi_1)+\cdots+G_n(\xi_n),~~z=\sum_j\xi_j$.} 
%{\Color{Maroon}FIG. 1}
%\end{center} 
\label{fig1-a}
\end{figure}

For the sake of completeness, let us illustrate how one applies
the result in (\ref{subp15b}) to obtain the results in (\ref{dglap19}).
We use the example of $P_{qq}$ for definiteness.
We apply the QCD exponentiation master formula 
embedded in eq.(\ref{subp15b})
%in our Appendix
%(see also Ref.~\cite{qcdexp}), following the
%analogous discussion then for QED in Refs.~\cite{jsw,jw},
to the gluon emission transition that
corresponds to $P_{qq}(z)$, i.e., to the squared amplitude for
$q\rightarrow q(z)+G(1-z)$ so that in the specialized
case already discussed above one replaces
everywhere the squared amplitudes for the $\bar{Q}'Q\rightarrow \bar{Q}'''Q''$
processes with those for the former one plus its $nG$ analogs 
with the attendant changes in the
phase space and kinematics dictated by the standard methods; this implies
that in eq.(53) of the first paper in Ref.~\cite{dglap} we
have from the application of the QCD aspect of eq.(\ref{subp15b}) 
the replacement (see Fig.~\ref{fig1-a})
\noindent
\begin{equation}
\begin{split}
P_{BA}&=P_{BA}^0\equiv\frac{1}{2}z(1-z)\overline{\underset{spins}{\sum}}~\frac{|V_{A\rightarrow B+C}|^2}{p_\perp^2}\\
&\Rightarrow\\
P_{BA}&=\frac{1}{2}z(1-z)\overline{\underset{spins}{\sum}}~\frac{|V_{A\rightarrow B+C}|^2}{p_\perp^2}z^{\gamma_q}F_{YFS}(\gamma_q)e^{\frac{1}{2}\delta_q} 
\end{split}
\label{expn1-dglp}
\end{equation}
where $A=q$, $B=G$, $C=q$ and $V_{A\rightarrow B+C}$ is the lowest order
amplitude
for $q\rightarrow G(z)+q(1-z)$, so that 
we get the un-normalized exponentiated result~\cite{irdglap1,irdglap2}
\begin{equation}
P_{qq}(z)= C_F F_{YFS}(\gamma_q)e^{\frac{1}{2}\delta_q}\frac{1+z^2}{1-z}(1-z)^{\gamma_q}.
\label{dglap8}
\end{equation} 
%where~\cite{irdglap1,irdglap2} 
%%\begin{equation}
%%\begin{split}
%\begin{align}
%\gamma_q &= C_F\frac{\alpha_s}{\pi}t=\frac{4C_F}{\beta_0}\\
%\delta_q&=\frac{\gamma_q}{2}+\frac{\alpha_sC_F}{\pi}(\frac{\pi^2}{3}-\frac{1}{2})
%%\end{split}
%\label{dglap9}
%\end{align}
%%\end{equation}
%and 
%\begin{equation}
%F_{YFS}(\gamma_q)=\frac{e^{-C_E\gamma_q}}{\Gamma(1+\gamma_q)}.
%\label{dglap10}
%\end{equation}
%Here, \[\beta_0=11-\frac{2}{3}n_f\], where $n_f$ is the number of
%active quark flavors, \[C_E=.5772\dots\] is Euler's constant
%and $\Gamma(w)$ is Euler's gamma function. The function
%$F_{YFS}(z)$ was already introduced by Yennie, Frautschi
%and Suura~\cite{yfs} in their analysis of the IR behavior
%of QED. 
We see immediately that
the exponentiation has removed the unintegrable IR divergence at $z=1$.
For reference, we note that we have in (\ref{dglap8}) resummed
the terms\footnote{Following the standard LEP Yellow Book~\cite{yellowbook} convention, we do not include the order of the first nonzero term in
counting the order of its higher order corrections.} ${\cal O}(\ln^k(1-z)t^{\ell}\alpha_s^n),~~n\ge\ell\ge k$, which originate
in the IR regime and which exponentiate. The important point is that
we have not dropped outright the terms that do not exponentiate
but have organized them into the residuals $\tilde{\bar\beta}_m$
in the analog of eq.(\ref{subp15b}).\par\indent

The application of eq.(\ref{subp15b}) to obtain eq.(\ref{dglap8})
proceeds as follows. First, the exponent in the exponential factor in front of the expression on the RHS of eq.(\ref{subp15b}) when restricted to QCD is readily seen to be 
%from 
%eq.(\ref{subp12})
, using the known results for the respective
real and virtual infrared functions from Refs.~\cite{irdglap1,irdglap2},
\begin{equation}
\begin{split}
{SUM}_{IR}(QCD)&=2\alpha_s \Re B_{QCD}+2\alpha_s\tilde B_{QCD}(\Kmax)\\
&=\frac{1}{2}\left(2 C_F\frac{\alpha_s}{\pi}t\ln{\frac{K_{max}}{E}}+ C_F\frac{\alpha_s}{2\pi}t+\frac{\alpha_sC_F}{\pi}(\frac{\pi^2}{3}-\frac{1}{2})\right)
\end{split}
\end{equation} 
where on the RHS of the last result we have already applied the DGLAP-CS
synthesization procedure as prescribed in Refs.~\cite{irdglap1,irdglap2} 
to remove the collinear
singularities, $\ln \Lambda^2_{QCD}/m_q^2 -1 $, in accordance with the standard
QCD factorization theorems~\cite{qcdfactorzn}. This means that, identifying
the LHS of eq.(\ref{subp15b}) as the sum over final states and average 
over initial states of the respective process divided by the incident
flux and replacing that incident flux by the respective initial state
density according to the standard methods for the
process $q\rightarrow q(1-z)+G(z)$, occurring in the context
of a hard scattering at scale $Q$ as it is for eq.(53) in the first paper
in Ref.~\cite{dglap}, the soft gluon effects for energy fraction
$<z\equiv K_{max}/E$
give the result, from eq.(\ref{subp15b}) restricted to QCD, that, working through to
the $\tilde{\bar\beta_1}$-level and using $q_2$ to represent the momentum
conservation via the other degrees of freedom for the attendant hard process, 
\begin{equation}
\begin{split}
\int\frac{\alpha_s(t)}{2\pi}P_{BA}dtdz&=e^{\rm SUM_{IR}(QCD)(z)}\int\{\tilde{\bar\beta}_0\int{d^4y\over(2\pi)^4}e^{\{iy\cdot(p_1-p_2)+\int^{k<K_{max}}{d^3k\over k}\tilde S_\rQCD(k)
\left[e^{-iy\cdot k}-1\right]\}}\\
&+ \int{d^3
k_1\over k_1}\tilde{\bar\beta}_1(k_1)\int{d^4y\over(2\pi)^4}e^{\{iy\cdot(p_1-p_2-k_1)+\int^{k<K_{max}}{d^3k\over k}\tilde S_\rQCD(k)
\left[e^{-iy\cdot k}-1\right]\}}\\
&+\cdots \}{d^3p_2\over p_2^{\,0}}{d^3q_2\over q_2^{\,0}} \\
&= e^{\rm SUM_{IR}(QCD)(z)}\int\{\tilde{\bar\beta}_0\int_{-\infty}^{\infty}{dy\over(2\pi)}e^{\{iy\cdot(E_1-E_2)+\int^{k<K_{max}}{d^3k\over k}\tilde S_\rQCD(k)
\left[e^{-iyk}-1\right]\}}\\
&+ \int{d^3
k_1\over k_1}\tilde{\bar\beta}_1(k_1)\int_{-\infty}^{\infty}{dy\over(2\pi)}e^{\{iy\cdot(E_1-E_2-k_1^0)+\int^{k<K_{max}}{d^3k\over k}\tilde S_\rQCD(k)
\left[e^{-iy\cdot k}-1\right]\}}\\
&+\cdots \}{d^3p_2\over p_2^{\,0}q_2^{\,0}} 
\end{split}
\label{expker1}
\end{equation}
where we set $E_i=p_i^0,~ i=1,2$ and the real infrared function $\tilde S_\rQCD(k)$ is known as well:
\begin{equation}
\tilde S_\rQCD(k)= -\frac{\alpha_s C_F}{8\pi^2}\left(\frac{p_1}{kp_1}-\frac{p_2}{kp_2}\right)^2|_{\text{DGLAP-CS synthesized}}
\label{realir}
\end{equation}
and we indicate as above that the DGLAP-CS synthesization
procedure as prescribed 
in Refs.~\cite{irdglap1,irdglap2} is to be applied to its evaluation to
remove its collinear singularities; we are using the 
kinematics of the first paper in Ref.~\cite{dglap} in their
computation of $P_{BA}(z)$ in their eq.(53), so that the 
relevant value of $k_\perp^2$ is indeed $Q^2$. It means that  
the computation can also be seen to correspond to computing the IR 
function for the standard t-channel kinematics and taking $\frac{1}{2}$ 
of the result to match the single line emission in $P_{Gq}$. 
The two integrals needed in (\ref{expker1}) were already
studied in Ref.~\cite{yfs}:
\begin{equation}
\begin{split}
I_{YFS}(zE,0)&=\int_{-\infty}^{\infty}\frac{dy}{2\pi}e^{[iy(zE)+\int^{k<zE}\frac{d^3k}{k}\tilde{S}_{QCD}(k)(e^{-iyk}-1)]}\\
&= F_{YFS}(\gamma_q)\frac{\gamma_q}{zE}\\
I_{YFS}(zE,k_1)&=\int_{-\infty}^{\infty}\frac{dy}{2\pi}e^{[iy(zE-k_1)+\int^{k<zE}\frac{d^3k}{k}\tilde{S}_{QCD}(k)(e^{-iyk}-1)]}\\
&=(\frac{zE}{zE-k_1})^{1-\gamma_q}I_{YFS}(zE,0)
\end{split}
\label{yfsintgls} 
\end{equation}
\par
When we introduce the results in (\ref{yfsintgls}) into (\ref{expker1})
we can identify the factor
\begin{equation}
\int\left(\tilde{\bar\beta}_0\frac{\gamma_q}{zE}+\int dk_1k_1d\Omega_1\tilde{\bar\beta}_1(k_1)(\frac{zE}{zE-k_1})^{1-\gamma_q}\frac{\gamma_q}{zE}\right)\frac{d^3p_2}{E_2q_2^{\,0}}=\int dt \frac{\alpha_s(t)}{2\pi}P_{BA}^0dz+{\cal O}(\alpha_s^2).
\label{1storder}
\end{equation}
where $P_{BA}^0$ is the unexponentiated result in the first line of
(\ref{expn1-dglp}). This leads us finally to the exponentiated 
result in the second line
of (\ref{expn1-dglp}) by elementary differentiation:
\begin{equation}
P_{BA}=P_{BA}^0z^{\gamma_q}F_{YFS}(\gamma_q)e^{\frac{1}{2}\delta_q}
\end{equation}
\par\indent
Let us stress the following. In this paper, we have retained for pedagogical
reasons the dominant terms in the resummation which we use for the kernels.
The result in the first line of (\ref{expker1}) is exact and can be used
to include all higher order resummation effects systematically as desired.
Moreover, we have taken a one-loop representation of $\alpha_s$ for illustration and have set it to a fixed-value on the RHS of (\ref{expker1}), 
so that, thereby, we are dropping further possible sub-leading higher order effects, again for reasons of pedagogy. It is straight forward to include these 
effects as well
%-- see Refs.~\cite{irdglap1,irdglap2} for the corresponding details.
-- see Refs.~\cite{irdglap1,irdglap2} for more discussion on this point.
Repeating the exhibited
resummation calculation for the other kernels leads to the
results in (\ref{dglap19}).
The latter results have now been implemented by 
MC methods, as we exhibit in the following sections.\par\indent

We stress that the improvement in Eq.\ (\ref{dglap19}) 
should be distinguished from 
the also important
resummation in parton density evolution for the ``$z\rightarrow 0$'' regime,
where Regge asymptotics obtain -- see for example Ref.~\cite{ermlv,guido}. This
latter improvement must also be taken into account 
for precision LHC predictions.\par\indent
 
%The results in Eq.\ (\ref{dglap19}) have now been implemented by 
%MC methods, as we exhibit in the following sections.\par

Let us now recall that already
a number of illustrative results and implications of the new 
kernels have been presented in Refs.~\cite{irdglap1,irdglap2,irdglap4-ichp}
which we summarize here as follows for the sake of completeness.
Firstly, we note that the connection to the higher order kernels in Refs.~\cite{high-ord-krnls} has been made in Ref.\ \cite{irdglap1}. This opens 
the way for the systematic improvement of the results presented herein.
Secondly, in the NS case, we find~\cite{irdglap1} that the $n=2$ moment
is modified by $\sim 5\%$ when evolved with Eq.\ (\ref{dglap19}) 
from $2$GeV to $100$GeV with $n_f=5$
and $\Lambda_{QCD}\cong 0.2$GeV, for illustration. This effect is thus relevant
to the expected precision of the HERA final data analysis~\cite{hera-dat}.
Thirdly, we have been able to use
Eq.\ (\ref{subp15b}) to resolve the violation~\cite{sac-no-go,cat1} 
of Bloch-Nordsieck cancellation in 
ISR(initial state radiation) 
at ${\cal O}(\alpha_s^2)$ for massive quarks~\cite{qmass-bw}.
This opens the way to include realistic quark masses as we introduce the
higher order EW corrections in the presence of higher order QCD corrections 
-- note that the radiation probability in QED at the hard scale $Q$ involves 
the logarithm $\ln(Q^2/m_q^2)$, and it will not do to set $m_q=0$ to analyze 
these effects in a fully exclusive, differential event-by-event calculation 
of the type that we are constructing. 
Fourthly, the threshold resummation implied by Eq.\ (\ref{subp15b}) for single $Z$
production at LHC shows a $0.3\%$ QED effect and agrees with known exact
results in QCD -- see Refs.~\cite{qced,baurall,exactqcd}. Fifthly, we have a 
new scheme~\cite{irdglap2} for precision LHC theory: in an obvious notation,
\begin{equation}
%\begin{split}
\sigma =\sum_{i,j}\int dx_1dx_2F_i(x_1)F_j(x_2)\hat\sigma(x_1x_2s)
       =\sum_{i,j}\int dx_1dx_2{F'}_i(x_1){F'}_j(x_2)\hat\sigma'(x_1x_2s),
%\end{split}
\label{sigscheme}
\end{equation}
where the primed quantities are associated with Eq.\ (\ref{dglap19}) in the
standard QCD factorization calculus. Sixthly, we have~\cite{qced} an attendant
shower/ME matching scheme, wherein, for example, in combining Eq.\ (\ref{subp15b})
with HERWIG~\cite{herwig}, PYTHIA~\cite{pythia}, MC@NLO~\cite{mcnlo}
or new shower MC's~\cite{skrzjad}, we may use either
$p_T$-matching
or shower-subtracted residuals\newline $\{\hat{\tilde{\bar\beta}}_{n,m}(k_1,\ldots,k_n;k'_1,\ldots,k'_m)\}$ to create a paradigm without double
counting that can be systematically improved order-by order in
perturbation theory -- see Refs.~\cite{qced}. 

The stage is set for the full MC implementation of our 
QED$\otimes$QCD resummation approach. We turn next to 
an important initial stage of this implementation -- that of the kernels in Eq.\ (\ref{dglap19}).

\section{MC Realization of IR-Improved DGLAP-CS  Theory}
In this section we describe the implementation of the 
new IR-improved kernels in the HERWIG6.5 environment, which results
in a new MC, which we denote by HERWIRI1.0, which stands for ``high energy radiation with IR improvement''\cite{bw-ms-priv-a}\par\indent

%footnote{We thank M. Seymour and B. Webber for discussion on this point.}. \par
Specifically, our approach can be summarized as follows.
We modify the kernels in the HERWIG6.5 module HWBRAN and in the attendant
 related modules~\cite{bw-ms-priv} with the following substitutions:
\begin{equation}\text{DGLAP-CS}\; P_{AB}  \Rightarrow \text{IR-I DGLAP-CS}\; P^{exp}_{AB}
\label{substitn}
\end{equation}\noindent
while leaving the hard processes alone for the moment. We have in 
progress~\cite{inprog}%% (SY,BFLW,MH,SM,SJ)
the inclusion of YFS synthesized electroweak  
modules from Refs.~\cite{jad-ward}%{\Color{Magenta}Jadach et al.  MC's} 
for
HERWIG6.5, HERWIG++~\cite{herpp} hard processes, as the
CTEQ~\cite{cteq} and MRST(MSTW)~\cite{mrst} best (after 2007) parton densities
do not include the precision electroweak higher order corrections that do enter in a 1\% precison tag budget for processes such as single heavy gauge boson production in the LHC environment~\cite{radcor-ew}. 
\par\indent

%%\end{itemize}
\def\beqa{\begin{eqnarray}}
\def\eeqa{\end{eqnarray}}
\def\beq{\begin{equation}}
\def\eeq{\end{equation}}
\def\non{\nonumber}
\def\no{\noindent }
For definiteness, let us illustrate the implementation by an example~\cite{bw-ann-rev,sjosback}, which for pedagogical reasons we will take as a simple leading
log shower component with a virtuality evolution variable, with the understanding that in HERWIG6.5 the shower development is angle ordered~\cite{bw-ann-rev} so that the evolution variable is actually $\sim E\theta$ where $\theta$ is the opening angle of the shower as defined in Ref.~\cite{bw-ann-rev} for a parton initial energy $E$. In this pedagogical example, which we take from Ref.~\cite{bw-ann-rev}, 
%\titbox{\Color{Maroon} Implementation Illustration}
the probability that no branching occurs above virtuality
cutoff $Q_0^2$ is  $\Delta_a(Q^2,Q_0^2)$ so that
%${\Color{Red}\Rightarrow}$
\beq \label{eq:splitprob}
d\Delta_a(t,Q_0^2) = \frac{-dt}{t}\Delta(t,Q_o^2)\sum_b\int dz\frac{\alpha_s}{2 \pi}P_{ba}(z),
\eeq
%\no
%${\Color{Red}\Rightarrow}$
which implies
\beq
\Delta_a(Q^2,Q_0^2)=\exp\left[ -\int_{Q_0^2}^{Q^2} \frac{dt}{t} \sum_b\int dz\frac{\alpha_s}{2 \pi}P_{ba}(z)\right].
\label{delta-a} 
\eeq
The attendant non-branching probability appearing in the evolution equation is
\beq
\Delta(Q^2,t) = \frac{\Delta_a(Q^2,Q_o^2)}{\Delta_a(t,Q_o^2)}, \quad t =k_a^2 \quad \text{the virtuality of gluon $a$}.
\eeq
The respective virtuality of parton $a$ is then generated with
\beq
\Delta_a(Q^2,t) = R,
\eeq
where $R$ is a random number uniformly distributed in $[0,1]$ .
With (note $\beta_0=b_0|_{n_c=3}$ here, where $n_c$ is the number of colors)
\beqa
\alpha_s(Q) = \frac{2 \pi}{b_0 \log\left(\frac{Q}{\Lambda}\right)},
\eeqa
we get for example 
\beqa
\int_0^1 dz \frac{\alpha_s(Q^2)}{2 \pi} P_{qG}(z)
&=& \frac{4\pi}{2 \pi b_0\ln\left(\frac{Q^2}{\Lambda^2}\right)}\int_0^1 dz \frac{1}{2}\left[ z^2+(1-z)^2\right] \non\\ 
&=& \frac{2}{3} \frac{1}{b_0\ln\left(\frac{Q^2}{\Lambda^2}\right)}.
\eeqa
so that the subsequent integration over $dt$ yields 
%${\Color{Red}\Rightarrow}$
\beqa
&&I=\int_{Q_0^2}^{Q^2}\frac{1}{3} \frac{dt}{t}\frac{2}{ b_0 \ln\left(\frac{t}{\Lambda^2}\right)} \non \\
%,\quad t=Q^2 \non \\
&=& \frac{2}{3b_0}\ln \ln \frac{t}{\Lambda^2}|_{Q_0^2}^{Q^2} \non \\
&=& \frac{2}{3 b_0}\left[\ln \left(\frac{\ln\left(\frac{Q^2}{\Lambda^2}\right)}{\ln\left(\frac{Q_0^2}{\Lambda^2}\right)}\right)\right].
\eeqa

Finally, introducing $I$ into Eq.\ (\ref{delta-a}) yields 
\beqa \label{DeltaQHerwig}
\Delta_a(Q^2,Q_0^2) &=& \exp \left[-\frac{2}{3 b_0}\ln \left(\frac{\ln\left(\frac{Q^2}{\Lambda^2}\right)}{\ln\left(\frac{Q_0^2}{\Lambda^2}\right)}\right)\right]\non\\
&=& \left[\frac{\ln\left(\frac{Q^2}{\Lambda^2}\right)}{\ln\left(\frac{Q_0^2}{\Lambda^2}\right)}\right]^{-\frac{2}{3b_0}}.
\eeqa
If we now let
$\Delta_a(Q^2,t)=R$, then
\beq 
\left[\frac{\ln\left(\frac{t}{\Lambda^2}\right)}{\ln\left(\frac{Q^2}{\Lambda^2}\right)}\right]^{\frac{2}{3b_0}} = R
\eeq
which implies
\beq
t = \Lambda^2 \left(\frac{Q^2}{\Lambda^2}\right)^{R^{\frac{3 b_0}{2}}}.
\label{t-herwig}
\eeq
Recall in HERWIG6.5~\cite{herwig} we have 
\beqa
b_0 &=& \left(\frac{11}{3}n_c - \frac{2}{3}n_f\right) \non \\ 
&=& \frac{1}{3}\left(11n_c - 10 \right), \quad n_f =5 \non \\
&\equiv& \frac{2}{3} {\tt BETAF}
\eeqa
where in the last line we used the notation in HERWIG6.5. 
%Note the
%simple relation $\beta_0=b_0|_{n_c=3}$ in our notations.
The momentum available after a  $q\bar{q}$ split in HERWIG6.5~\cite{herwig} 
is given by
\beq
{\tt QQBAR} = {\tt QCDL3} \left(\frac{\tt QLST}{\tt QCDL3}\right)^{R^{\tt BETAF}},
\eeq
in complete agreement with Eq.\ (\ref{t-herwig}) when we note the
identifications $t={\tt QQBAR}^2,\;\Lambda\equiv {\tt QCDL3},
\; Q\equiv {\tt QLST}$.

The leading log exercise leads to the same algebraic relationship that
HERWIG6.5 has between {\tt QQBAR} and {\tt QLST} but 
we stress that in HERWIG6.5
these quantities are the angle-ordered counterparts of the virtualities 
we used in our example, so that the shower is angle-ordered.\par\indent

\begin{figure}[h]
\begin{center}
%\scalebox{0.99}{\includegraphics[angle=90]{delta.eps}}
\includegraphics[height=150mm]{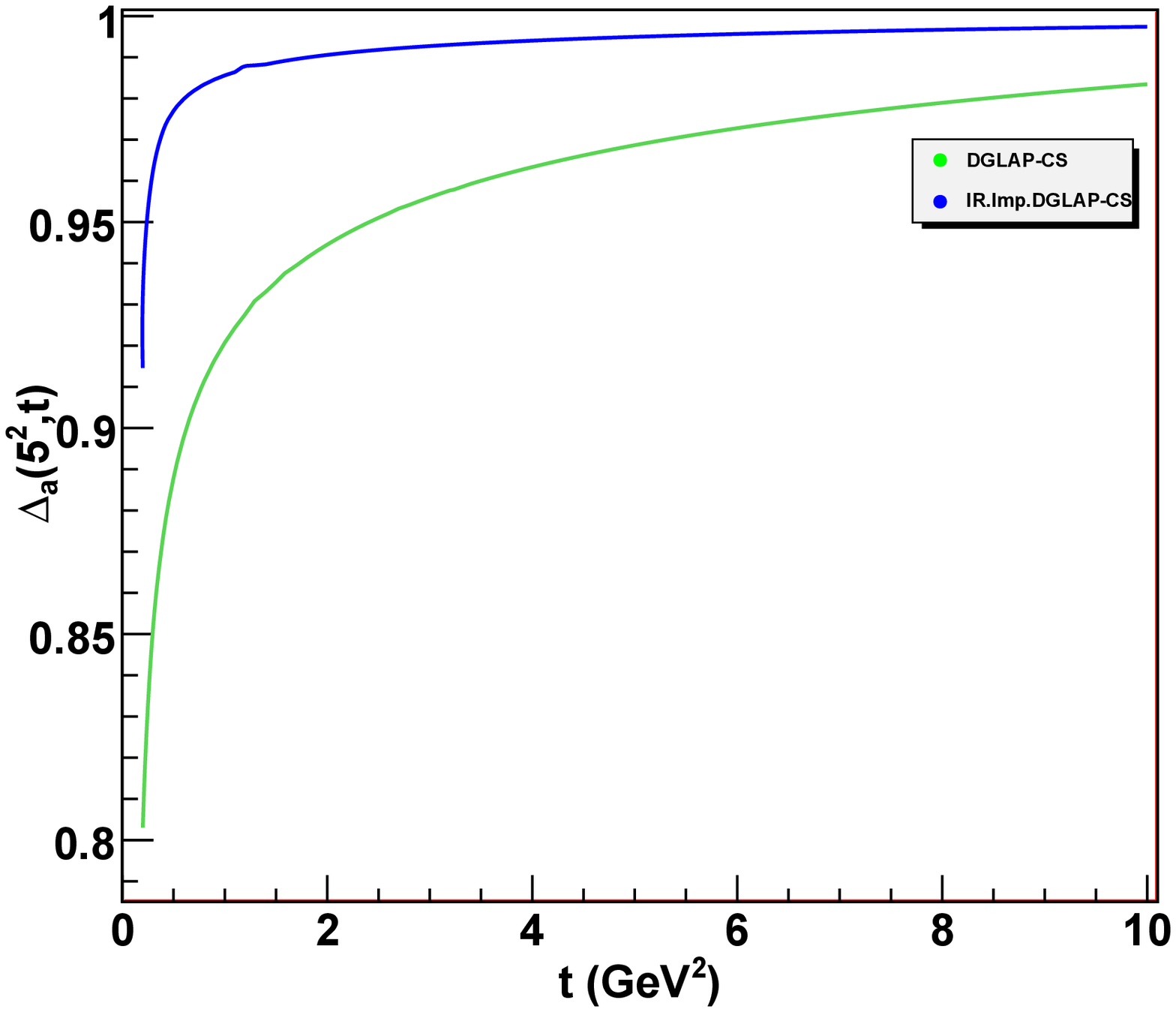}
\end{center}
\caption{ Graph of $\Delta_a(Q^2,t)$ for the DGLAP-CS and IR-Improved DGLAP-CS kernels Eqs. (\ref{DeltaQHerwig}, \ref{DeltaQWard}). Q$^2$ is a typical virtuality close to the squared scale of the hard sub-process 
-- here we use $Q^2=25$GeV$^2$ for illustration.} 
\label{iri-vs-cs}
\end{figure}

Let us now repeat the above calculation for the IR-Improved kernels in 
Eq.\ (\ref{dglap19}). We have
\beq
P_{qG}^{\exp}(z) = \FYFS(\gamma_G)e^{\delta_G/2}\frac{1}{2} 
                    \biggl[ z^2(1-z)^{\gamma_G} + (1-z)^2z^{\gamma_G} \biggr]
\eeq
so that
\beq
\int_0^1 dz \frac{\alpha_s\left(Q^2\right)}{2 \pi} P_{qG}(z)^{\exp}
 = \frac{4\FYFS(\gamma_G)e^{\delta_G/2} }{b_0 \ln\left(\frac{Q^2}{\Lambda^2}\right)\left(\gamma_G+1\right)\left(\gamma_G+2\right)\left(\gamma_G+3\right)}.
\eeq
This leads to the following integral over $dt$:
\beqa
&&I=\int_{Q_0^2}^{Q^2}\frac{dt}{t} \frac{4\FYFS(\gamma_G)e^{\delta_G/2} }{b_0 \ln\left(\frac{t}{\Lambda^2}\right)\left(\gamma_G+1\right)\left(\gamma_G+2\right)\left(\gamma_G+3\right)} \non\\
%,\quad t=Q^2 \non \\ 
&=&\frac{4 \FYFS(\gamma_G)e^{\gamma_G/4}}{b_0\left(\gamma_G+1\right)\left(\gamma_G+2\right)\left(\gamma_G+3\right)}
\Ei\left(1,\frac{8.369604402}{b_0\ln\left(\frac{t}{\Lambda^2}\right)}\right) \Bigg\vert_{Q_0^2}^{Q^2}.
\eeqa
%Here we have used
%\beq
%\delta_G=\frac{\gamma_G}{2} + \frac{\alpha_s C_G}{\pi}\left(\frac{\pi^2}{3} - \frac{1}{2}\right),
%\eeq
%with $C_G=3$ the gluon quadratic Casimir invariant.
We finally get the IR-improved formula
\beq \label{DeltaQWard}
\Delta_a(Q^2,t) = \exp\left[-\left(F\left(Q^2\right)-F\left(t\right)\right)\right],
\eeq
where
\beq
F(Q^2) = \frac{4 \FYFS(\gamma_G)e^{\gamma_G/4}}{b_0\left(\gamma_G+1\right)\left(\gamma_G+2\right)\left(\gamma_G+3\right)}
\Ei\left(1,\frac{8.369604402}{b_0\ln\left(\frac{Q^2}{\Lambda^2}\right)}\right),
\eeq
and $\Ei$  is the exponential integral function.
In Fig.~\ref{iri-vs-cs} we show the difference between the two results for 
$\Delta_a(Q^2,t)$. We see that they agree within a few percent except for the softer values of $t$, as expected. We look forward to determining definitively
whether the experimental data prefer one over the other. This detailed study will appear elsewhere~\cite{elswh} but we begin the discussion below with a view on recent FNAL data. 
Again, we note that the comparison in Fig.~\ref{iri-vs-cs} is carried out at the leading log virtuality level, but the sub-leading effects suppressed in this 
discussion will not change our general conclusions drawn therefrom.

For further illustration, we note that
for the $q\rightarrow qG$ branching process in HERWIG6.5~\cite{herwig}, we have therein the implementation of the usual DGLAP-CS kernel as follows:

\begin{minipage}[c]{0.75\linewidth}
\begin{verbatim}
      WMIN = MIN(ZMIN*(1. -ZMIN), ZMAX*(1.-ZMAX))
      ETEST = (1. + ZMAX**2) * HWUALF(5-SUDORD*2, QNOW*WMIN)
      ZRAT = ZMAX/ZMIN
30    Z1 = ZMIN * ZRAT**HWRGEN(0)
      Z2 = 1. - Z1
      PGQW = (1. + Z2*Z2)
      ZTEST = PGQW * HWUALF(5-SUDORD*2, QNOW*Z1*Z2)
      IF (ZTEST .LT. ETEST*HWRGEN(1)) GOTO 30
      ...

\end{verbatim} 
\end{minipage}\begin{minipage}[c]{0.25\linewidth}
\begin{equation}
\hfill
\label{exhwb1}
\end{equation}
\end{minipage}
%\begin{align}
%\;\;\;\;\;&WMIN=MIN(ZMIN*(1. -ZMIN),ZMAX*(1.-ZMAX))\nonumber\\
%\;\;\;\;\;&ETEST=(1.+ZMAX**2)*HWUALF(5-SUDORD*2,QNOW*WMIN)\nonumber\\
%\;\;\;\;\;&ZRAT=ZMAX/ZMIN\nonumber\\
%\;\;\;30\;&Z1=ZMIN*ZRAT**HWRGEN(0)\nonumber\\
%\;\;\;\;\;&Z2=1.-Z1\nonumber\\
%\;\;\;\;\;&PGQW = (1.+Z2*Z2)\nonumber\\
%\;\;\;\;\;&ZTEST=PGQW*HWUALF(5-SUDORD*2,QNOW*Z1*Z2)\nonumber\\
%\;\;\;\;\;&IF(ZTEST.LT.ETEST*HWRGEN(1)) GOTO\; 30\nonumber\\
%\;\;\;\;\;&\ldots
%\label{exhwb1}
%\end{align}
where the branching of $q$ to $G$ at $z=${\tt Z1} occurs in the interval from
{\tt ZMIN} to {\tt ZMAX} set by the inputs to the program and the current 
value of the virtuality {\tt QNOW}, {\tt HWUALF} is the respective function 
for $\alpha_s$ in the program and {\tt HWRGEN(J)} are uniformly 
distributed random numbers on the interval from 0 to 1. It is seen that
Eq.\ (\ref{exhwb1}) is a standard MC realization of the unexponentiated DGLAP-CS kernel via
\begin{equation}
\alpha_s(Qz(1-z))P_{Gq}(z)=\alpha_s(Qz(1-z))\frac{1+(1-z)^2}{z}
\end{equation}
where the normalization is set by the usual conservation of probability.
To realize this with the IR-improved kernel, we make the replacement
of the code in Eq.\ (\ref{exhwb1}) with the lines 
\begin{minipage}[c]{0.75\linewidth}
\begin{verbatim}
      NUMFLAV = 5
      B0 = 11. - 2./3.*NUMFLAV
      L = 16./(3.*B0)
      DELTAQ = L/2 + HWUALF(5-SUDORD*2, QNOW*WMIN)*1.184056810
      ETEST = (1. + ZMAX**2) * HWUALF(5-SUDORD*2, QNOW*WMIN)
            * EXP(0.5*DELTAQ) * FYFSQ(NUMFLAV-1) * ZMAX**L
      ZRAT = ZMAX/ZMIN
30    Z1 = ZMIN * ZRAT**HWRGEN(0)
      Z2 = 1. - Z1
      DELTAQ = L/2 + HWUALF(5-SUDORD*2, QNOW*Z1*Z2)*1.184056810
      PGQW = (1. + Z2*Z2) * EXP(0.5*DELTAQ) * FYFSQ(NUMFLAV-1)
           * Z1**L
      ZTEST = PGQW * HWUALF(5-SUDORD*2, QNOW*Z1*Z2)
      IF (ZTEST .LT. ETEST*HWRGEN(1)) GOTO 30
      ...

\end{verbatim} 
\end{minipage}\begin{minipage}[c]{0.25\linewidth}
\begin{equation}
\hfill
\label{exhwb2}
\end{equation}
\end{minipage}
%\begin{align}
%\;\;\;\;\;\;&NUMFLAV=5\nonumber\\
%\;\;\;\;\;\;&b0 = 11. -2./3*NUMFLAV\nonumber\\
%\;\;\;\;\;\;&l = 16./(3*b0)\nonumber\\
%\;\;\;\;\;\;&DELTAq =l/2+HWUALF(5-SUDORD*2,QNOW*WMIN)*1.184056810\nonumber\\
%\;\;\;\;\;\;&ETEST=(1.+ZMAX**2)*HWUALF(5-SUDORD*2,QNOW*WMIN)\nonumber\\
%\;\;\;\;\&\;& \qquad     *exp(.5*DELTAq)*FYFSq(NUMFLAV-1)*ZMAX**l\nonumber\\
%\;\;\;\;\;\;&ZRAT=ZMAX/ZMIN\nonumber\\
%\;\;30\;\;\;&Z1=ZMIN*ZRAT**HWRGEN(0)\nonumber\\
%\;\;\;\;\;\;&Z2=1.-Z1\nonumber\\
%\;\;\;\;\;\;&DELTAq =l/2+HWUALF(5-SUDORD*2,QNOW*Z1*Z2)*1.184056810\nonumber\\
%\;\;\;\;\;\;&PGQW = (1. + Z2*Z2)*exp(.5*DELTAq)*FYFSq(NUMFLAV-1)\nonumber\\
%\;\;\;\;\&\;& \qquad     *Z1**l\nonumber\\
%\;\;\;\;\;\;&ZTEST=PGQW*HWUALF(5-SUDORD*2,QNOW*Z1*Z2)\nonumber\\
%\;\;\;\;\;\;&IF (ZTEST.LT.ETEST*HWRGEN(1)) GOTO\; 30\nonumber\\
%\;\;\;\;\;\;&\ldots
%\label{exhwb2}
%\end{align}
so that with the identifications $\gamma_q\equiv {\tt L},\; \delta_q\equiv 
{\tt DELTAQ},\; \FYFS(\gamma_q)\equiv {\tt FYFSQ(NUMFLAV-1)}$, we see 
that Eq.\ (\ref{exhwb2}) realizes the IR-improved DGLAP-CS kernel 
$P^{\exp}_{Gq}(z)$ via $\alpha_s(Qz(1-z))P^{\exp}_{Gq}(z)$ with the 
normalization again set by probability conservation. 
\par\indent

Continuing in this way, 
we have  carried out the corresponding changes for all of the kernels 
in Eq.\ (\ref{dglap19}) in the HERWIG6.5 environment, with its angle-ordered 
showers, resulting in the new MC, HERWIRI1.0(31),
in which the ISR parton showers have IR-improvement as given by
the kernels in Eq.\ (\ref{substitn}).
In the original 
release of the program, v. 1.0, 
we stated that the time-like parton showers had 
been completely IR-improved in a way that suggested the space-like parton 
showers had not yet been IR-improved at all. 
In the subsequent release, v. 1.02, the part of the 
space-like parton showers without IR-improvement associated with 
HERWIG6.5's space-like module HWSGQQ for the space-like branching process
$G\rightarrow q\bar{q}$, a process which is not IR divergent and which is, 
in any case, a sub-dominant part of the shower, was IR-improved. In the release
in version 1.031 the final missing IR-improvement in the space-like
module HWSFBR\footnote{We thank M. Seymour and B. Webber for discussion on this point.} has been implemented.
The IR-improvement of the module HWSGQQ in the release HERWIRI1.02 produces
a small effect, as these considerations suggest: we see effects at a 
level comparable to the errors on the MC data 
in our plots when going from version 1.0 to 1.02. In going from version 1.0 to 1.031, we do some significant effects in the soft $p_T$ regime so that 
we recommend the use of version 1.031 in comparison with data, as we illustrate
presently.
\par\indent

We now illustrate some of the results we have obtained in comparing 
ISR showers in HERWIG6.5 and with those in HERWIRI1.0(31) at LHC
and at FNAL
energies, where some comparison with real data is also featured at the FNAL
energy. Specifically, we compare the $z$-distributions, $p_T$-distributions, etc., that result from the IR-improved and usual
DGLAP-CS showers in what follows~\cite{similar}.\par\indent

\begin{figure}[h]
\begin{center}
\includegraphics[height=200mm,angle=90]{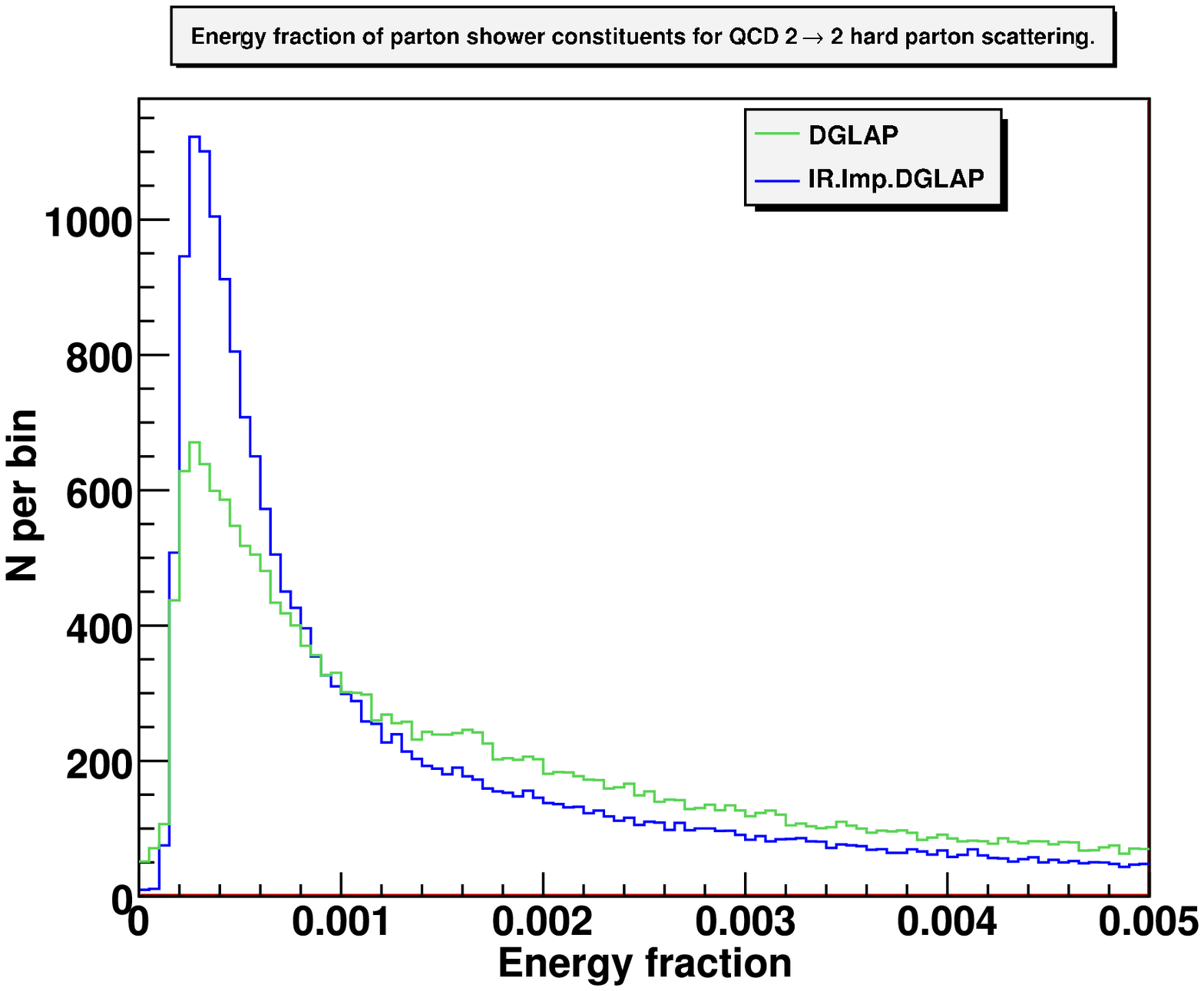}
\end{center}
\caption{ The $z$-distribution(ISR parton energy fraction) shower comparison in HERWIG6.5.}
\label{fighw1}
\end{figure}

\begin{figure}[h]
\begin{center}
\includegraphics[height=150mm]{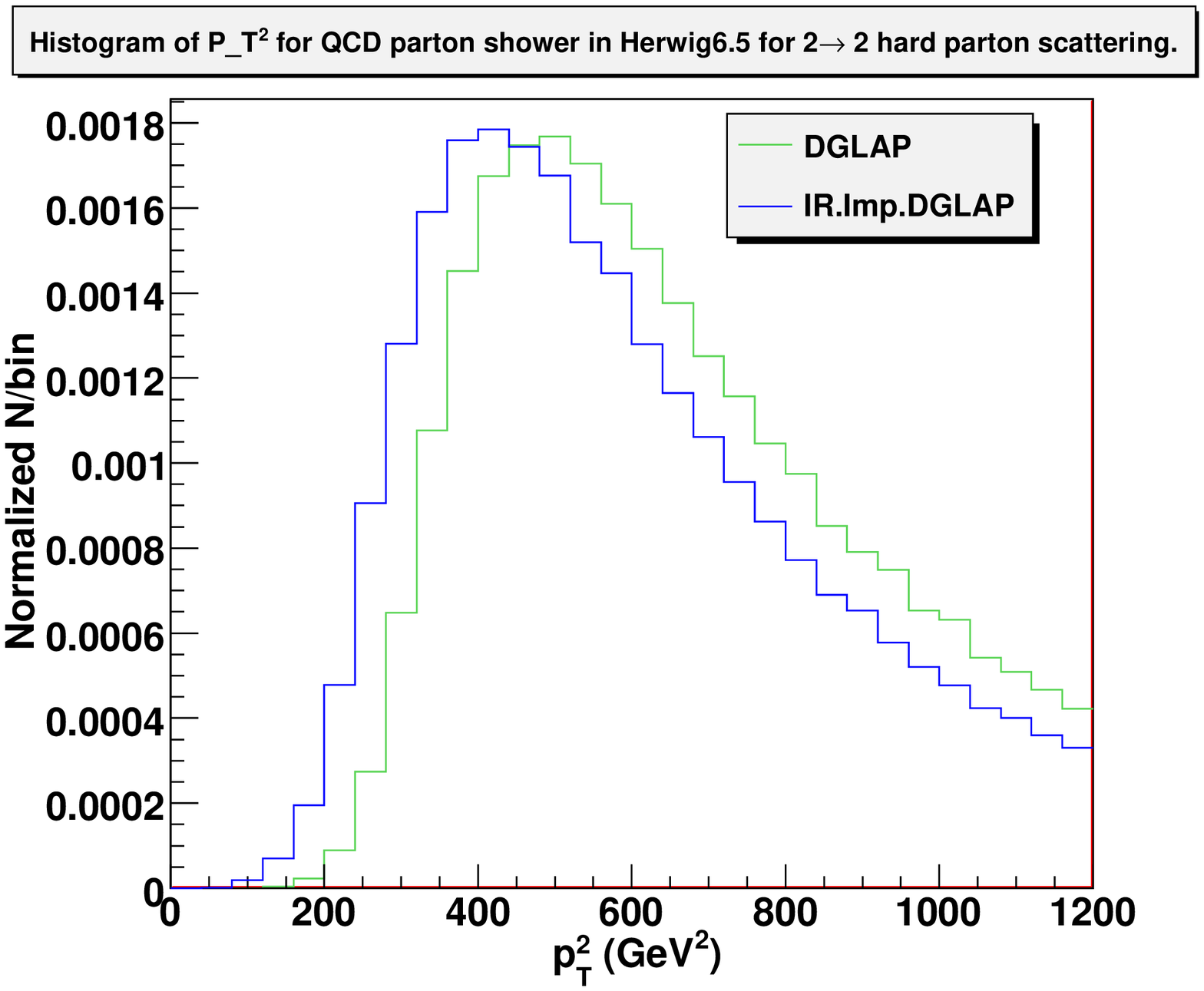}
\end{center}
\caption{The  $p_T^2$-distribution (ISR parton) shower comparison in HERWIG6.5.}
\label{fighw2}
\end{figure}

%\footnote{Note that similar results for 
%PYTHIA and MC@NLO are in progress.}.\par
First, for the generic 2$\rightarrow$2 hard processes at LHC energies (14 TeV) we get the comparison shown Figs.~\ref{fighw1}, \ref{fighw2} for the respective ISR $z$-distribution and $p_T^2$ distribution at the parton level. 
Here, there are no cuts placed on the MC data and we define $z$ as
$z=E_{\text{parton}}/E_{\text{beam}}$ where $E_{\text{beam}}$ is the cms beam energy and $E_{\text{parton}}$ is the respective parton energy in the cms system. The two quantities $z$ and  $p_T^2$ for partons are of course not directly observable but their distributions show the softening of the IR divergence as we expect.

%%\ref{fighw1},\ref{fighw2},\ref{fighw3}:
\begin{figure}[h]
\begin{center}
\includegraphics[width=150mm]{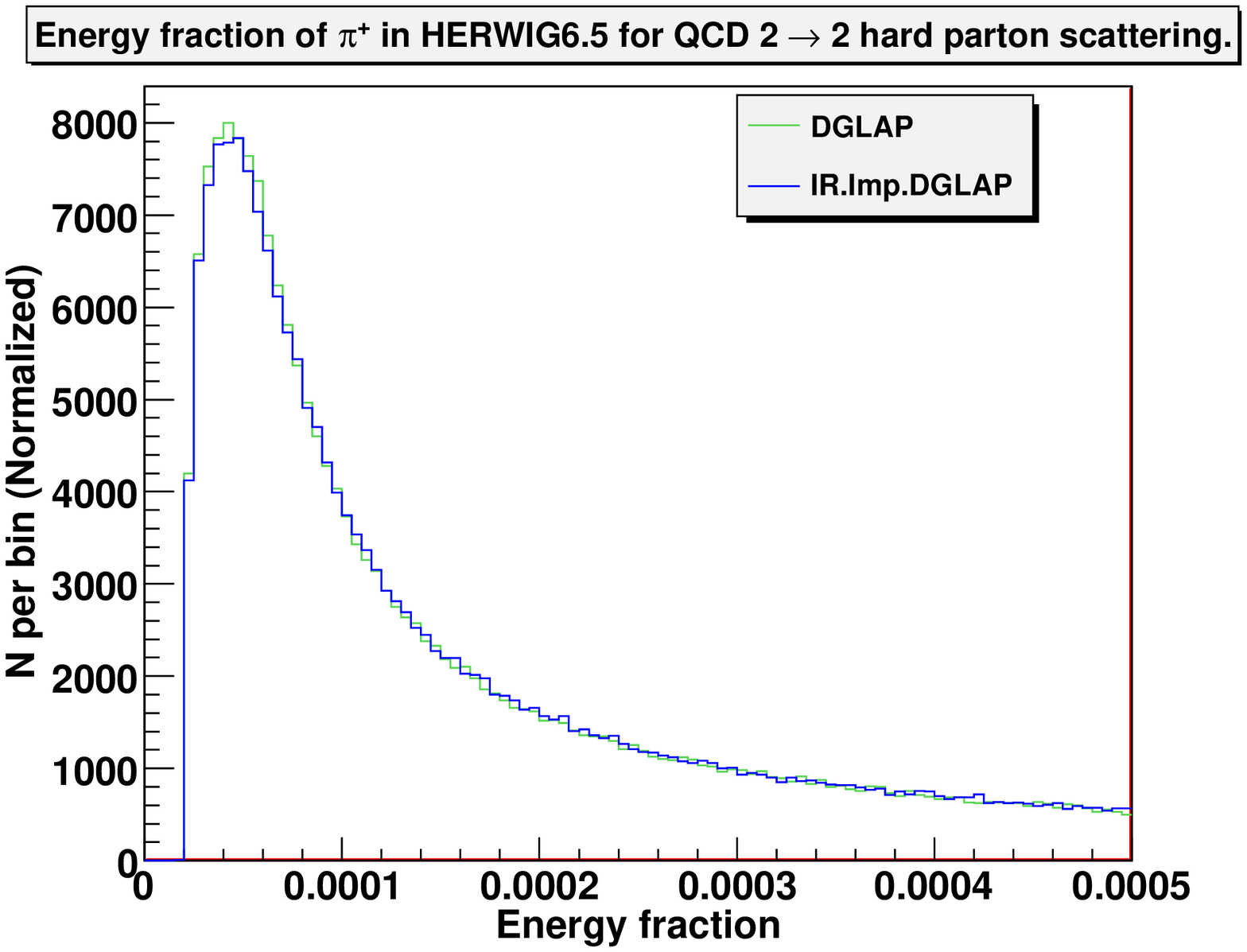}
\end{center}
\caption{The $\pi^+$ energy fraction distribution shower comparison in HERWIG6.5.}
\label{fighw3}
\end{figure}

\begin{figure}[h]
\begin{center}
\includegraphics[height=150mm]{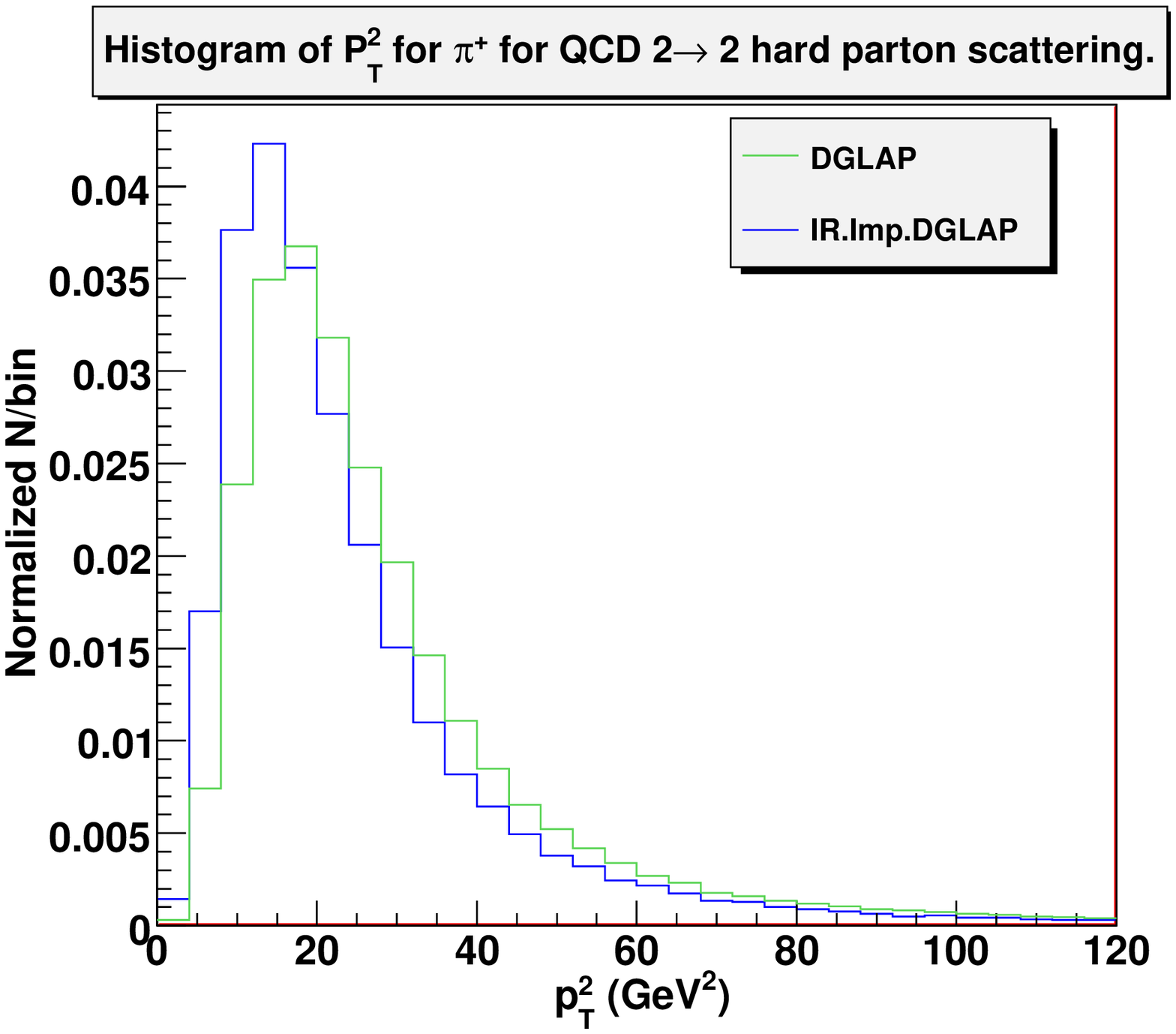}
\end{center}
\caption{\protect The $\pi^+$ $p_T^2$-distribution shower comparison in HERWIG6.5.}
\label{fighw4}
\end{figure}

%\ref{fighw1},\ref{fighw2},\ref{fighw3}:
Turning next to the similar quantities for the $\pi^+$ production in the 
generic 2$\rightarrow$2 hard processes at LHC, we see in Figs.~\ref{fighw3}, \ref{fighw4} that spectra in the former are similar and spectra in the latter are again softer in the IR-improved case. These spectra of course would be 
subject to some ``tuning'' in a real experiment and we await with anticipation the outcome of such 
an effort in comparison to LHC data.\par\indent

\begin{figure}[h]
\begin{center}
\includegraphics[width=150mm]{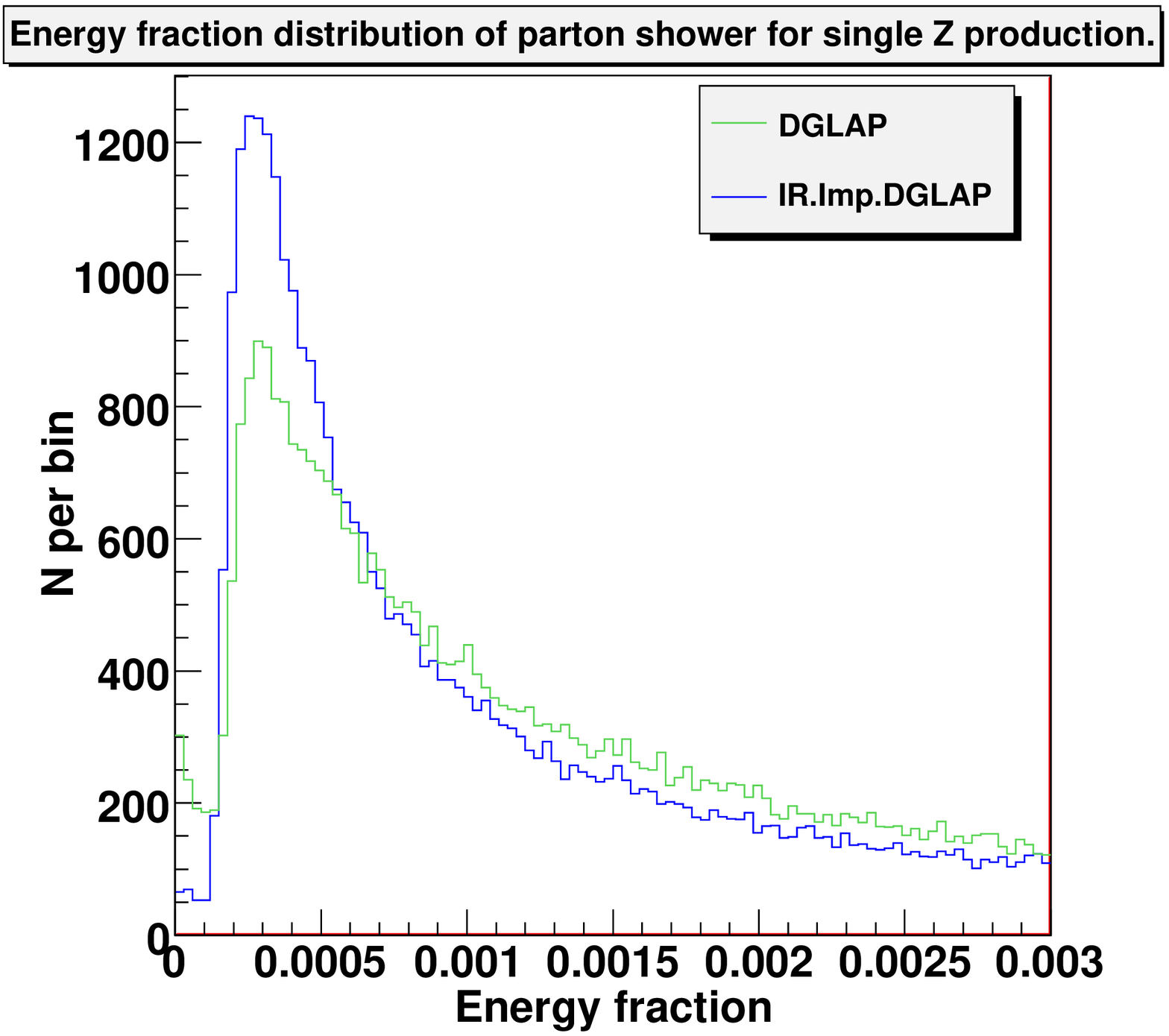}
\end{center}
\caption{The $z$-distribution(ISR parton energy fraction) shower comparison in HERWIG6.5.}
\label{fighw5}
\end{figure}

\begin{figure}[h]
\begin{center}
\includegraphics[height=150mm]{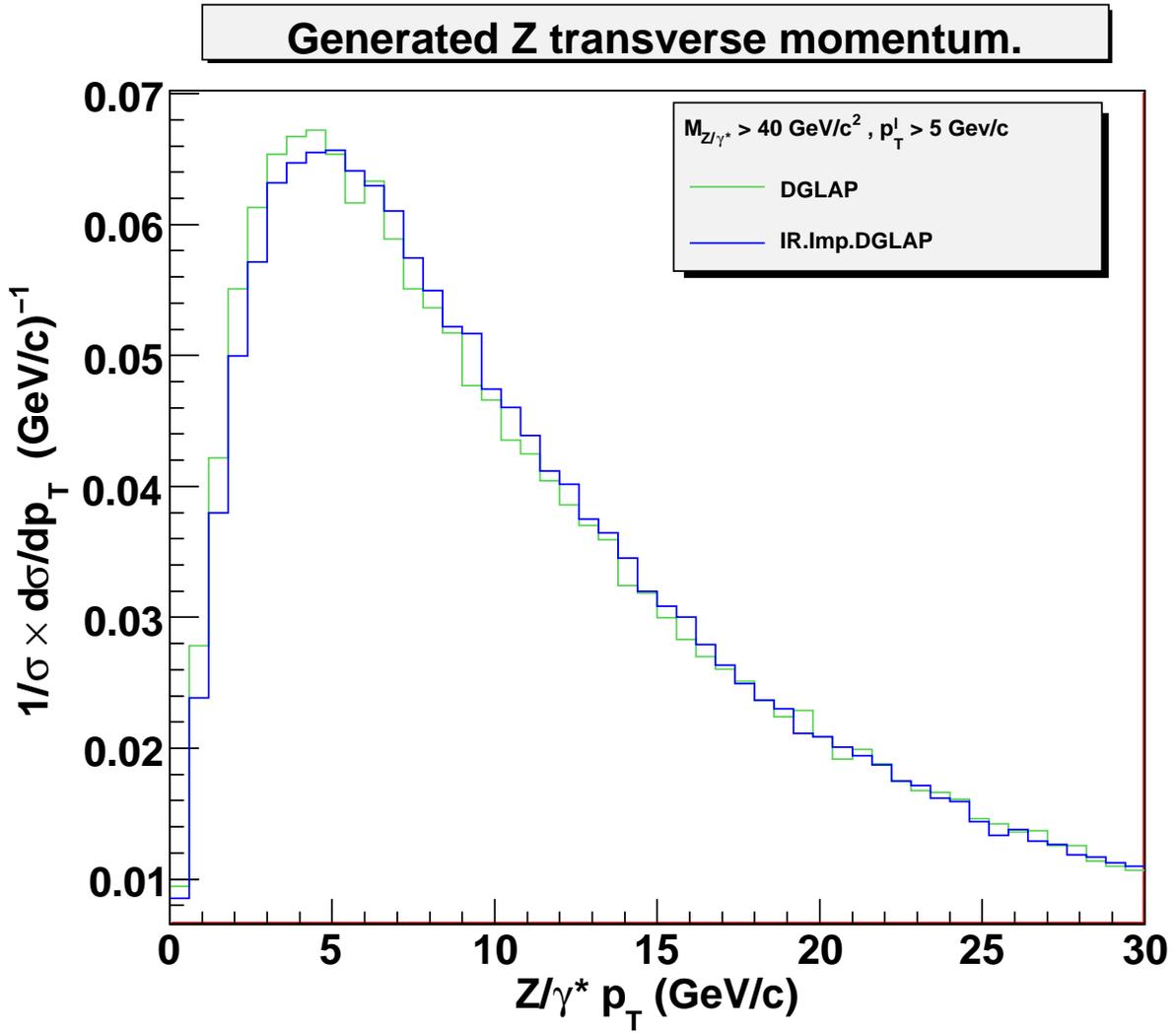}
\end{center}
\caption{The $Z$ p$_T$-distribution(ISR parton shower effect) comparison in HERWIG6.5.}
\label{fighw6}
\end{figure}

\begin{figure}[h]
\begin{center}
\includegraphics[height=150mm]{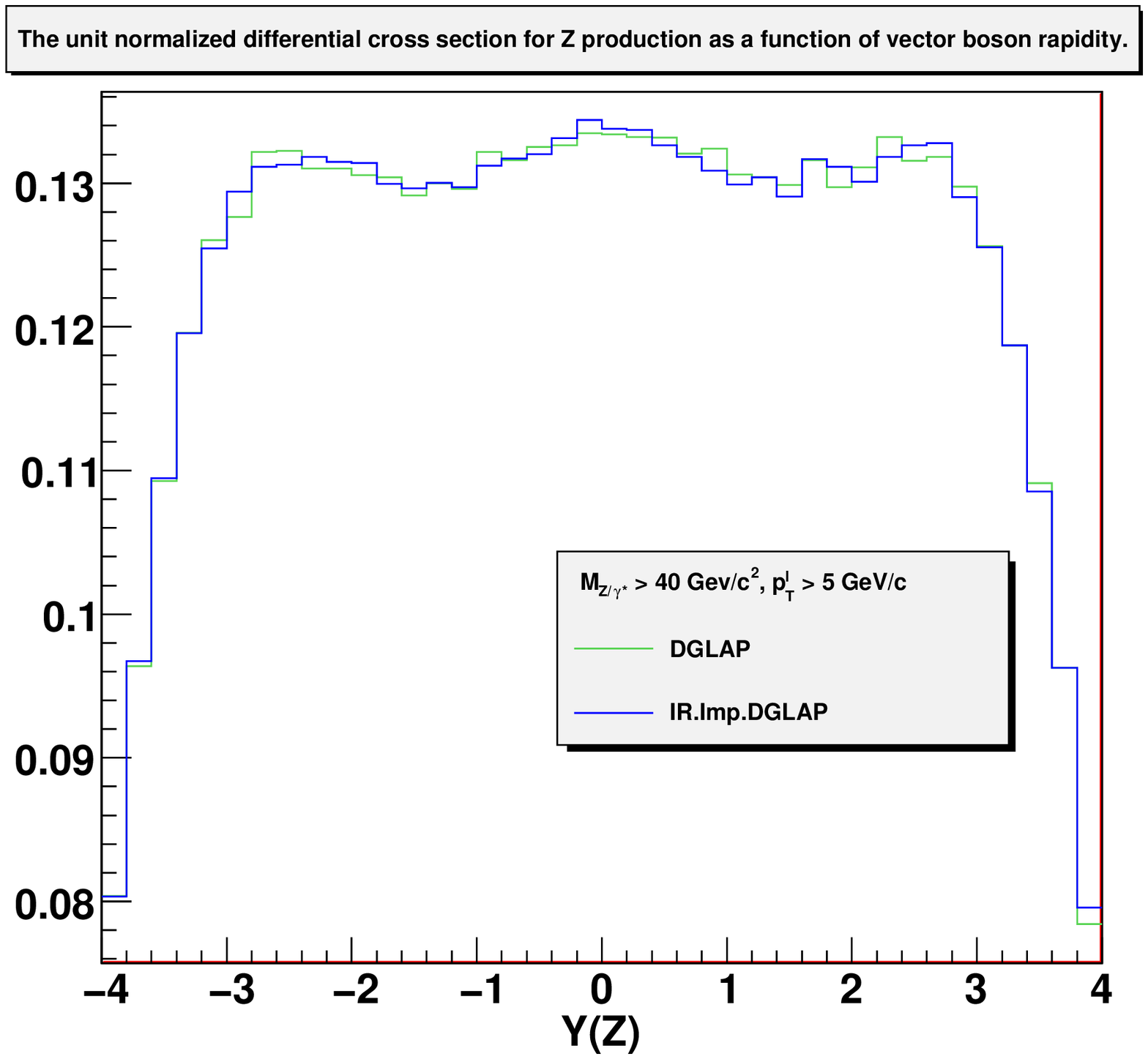}
\end{center}
\caption{The $Z$ rapidity-distribution(ISR parton shower) comparison in HERWIG6.5.}
\label{fighw7}
\end{figure}

We turn next to the luminosity process of single $Z$ production at the LHC, 
where in Figs.~\ref{fighw5},\ref{fighw6},\ref{fighw7} we show respectively 
the ISR parton energy fraction distribution, the $Z$ p$_T$ distribution, 
and the $Z$ rapidity distribution
with cuts on the acceptance as $40\text{GeV}<M_Z,\; p^\ell_T>5\text{GeV}$
%,\; |\eta_\ell|<50$ 
for $Z\rightarrow \mu^+\mu^-$ -- all lepton rapidities are
included. For the energy fraction distribution and the p$_T$ 
distributions we again see softer spectra in the former and similar spectra in the latter in the IR-improved case. 
For the rapidity plot, we see the migration of some events to the 
higher values of $Y$, which is not 
inconsistent with a softer spectrum for the IR-improved case.
One might wonder why we show the $Z$ rapidity here as the soft gluons
which we study only have an indirect affect on it via momentum conservation?
But, this means that the rapidity predicted by the IR-improved showers
should be close to that predicted by the un-improved showers 
and we show this cross-check is
indeed fulfilled in our plots. To understand why one has the migration to 
higher values of rapidity 
in the IR-improved spectra, recall the IR-improved spectra
move the radiated partons to softer values of $z$ and this means the
produced $Z$'s have harder values of energy for given $p_T$
as the $p_T$ spectra are similar, and this in turn means
these $Z$'s have higher values of the rapidity variable. 
We look forward to the confrontation with experiment,
where again we stress that in a real experiment, a certain amount of ``tuning'' will affect these results. We note for example that the difference between the
spectra in Fig.~\ref{fighw6}, while it is interesting, is well within
the range that could be tuned away by varying the amount of
intrinsic transverse momentum of partons in the proton. The question will always be which set of distributions gives a better $\chi^2$ per degree of freedom.\par\indent

%\item Single Z-production at LHC
\begin{figure*}[t]
%\begin{center}
%%%%\epsfig{file=fig01.ps}
%\epsfig{file=en_sgam-200.eps,width=140mm,height=130mm}
%\end{center}
%\vspace{ -8mm}
%\baselineskip=7mm
\centering
\setlength{\unitlength}{0.1mm}
%%%%%%%%%%%%%%%%%%%%%%%%%%%%%%%%%%
%%%\begin{picture}(1600, 1540)
\begin{picture}(1600, 950)
%%\put( 450, 1530){\makebox(0,0)[cb]{\bf (a)} }
%%\put(1230, 1530){\makebox(0,0)[cb]{\bf (b)} }
%%\put(   0, 870){\makebox(0,0)[lb]{\epsfig{file=ef.ps,angle=270,
%%                                        width=80mm}}}
%%\put( 800, 870){\makebox(0,0)[lb]{\epsfig{file=hwptc.ps,angle=270,
%%                                        width=80mm}}}
%%%\put( 250, 660){\makebox(0,0)[cb]{\bf (a)} }
%%%\put(800, 660){\makebox(0,0)[cb]{\bf (b)} }
%%%\put(   0, 100){\makebox(0,0)[lb]{\epsfig{file=vgcut.eps,angle=90,
%%%                                        width=55mm}}}
%%%\put( 550, 100){\makebox(0,0)[lb]{\epsfig{file=vgcutir.eps,angle=90,
%%%                                        width=55mm}}}
\put( 350, 900){\makebox(0,0)[cb]{\bf (a)} }
\put(1330, 900){\makebox(0,0)[cb]{\bf (b)} }
\put(   -150, 0){\makebox(0,0)[lb]
%{\includegraphics[width=75mm,angle=90]{endglap.ps}}}
{\epsfig{file=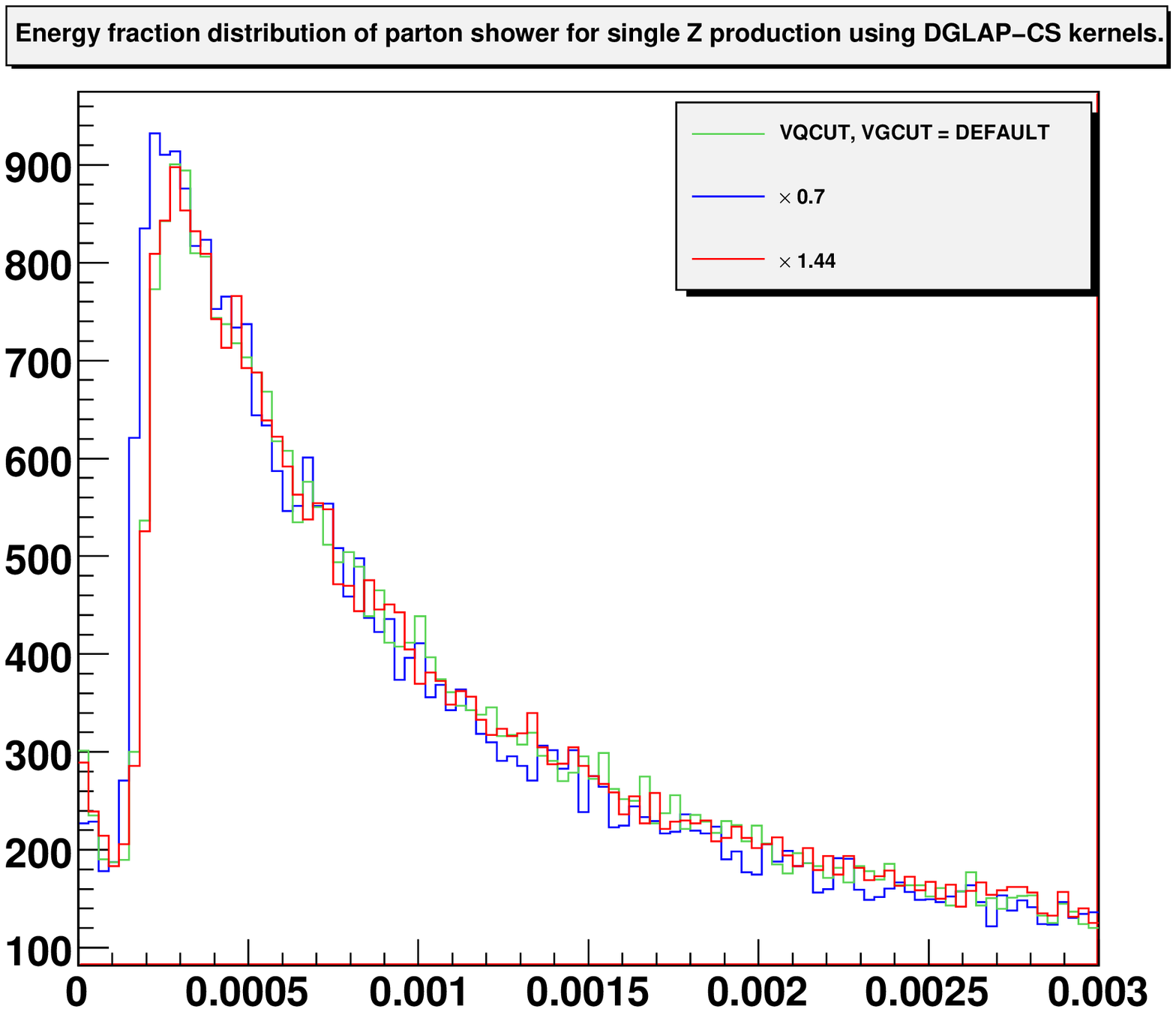,
                                        width=100mm}}}
\put( 860, 0){\makebox(0,0)[lb]
%{\includegraphics[width=70mm,angle=270]{endglpirspwrd.eps}}}
{\epsfig{file=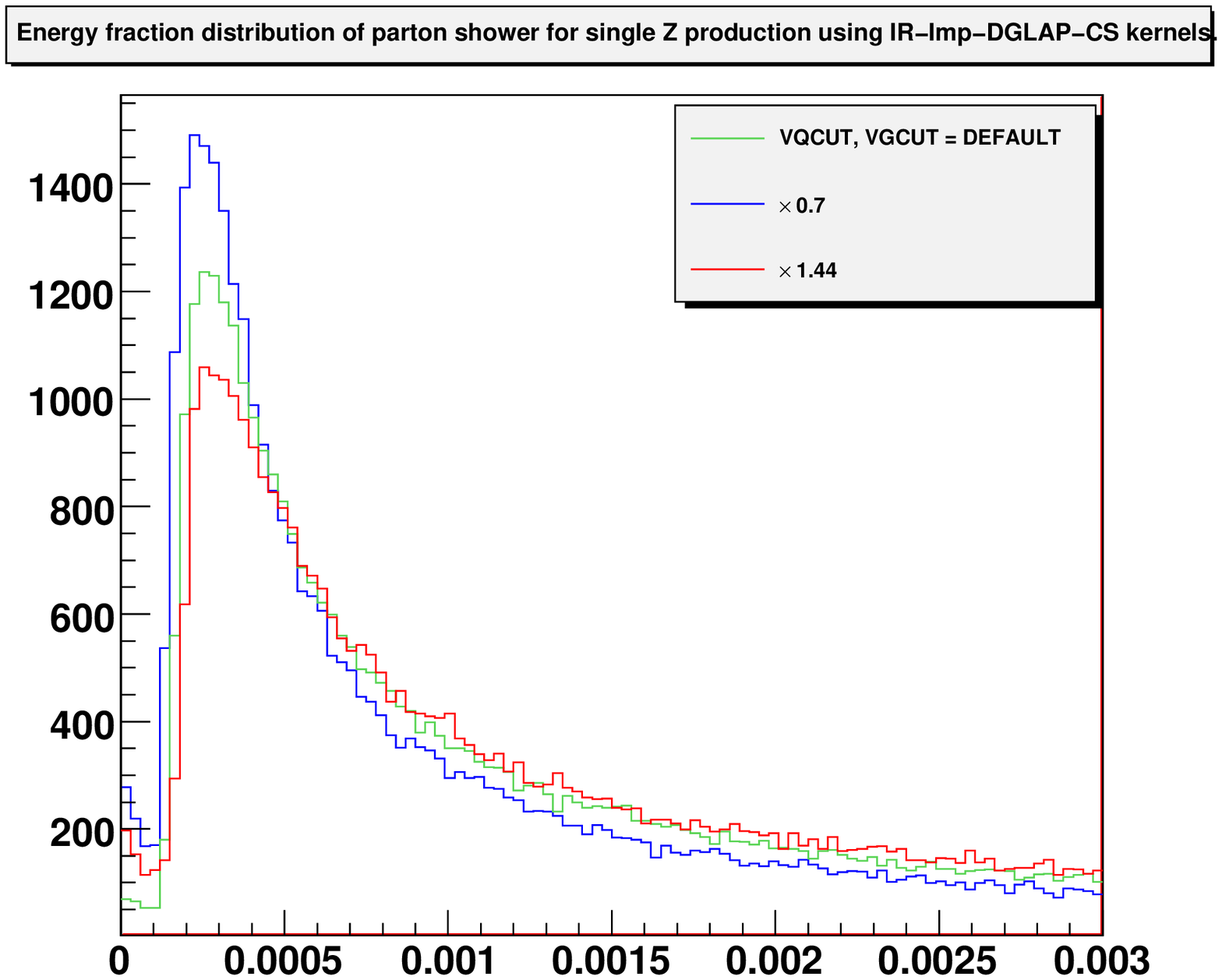,
                                        width=100mm}}}
\end{picture}
%\vspace{ -1.5mm}
\caption{IR-cut-off sensitivity in z-distributions of the ISR parton energy fraction: (a), DGLAP-CS 
(b), IR-I DGLAP-CS -- for the single $Z$ hard
sub-process in HERWIG-6.5 environment.
%\centerline{\Color{PineGreen}COMPARISON WITH DATA IMMINENT.} 
}
\label{fighw8}
\end{figure*}

Finally, we turn the issue of the IR-cut-off in HERWIG6.5. In HERWIG6.5,
there is are IR-cut-off parameters used to separate real and virtual effects
and necessitated by the +-function representation of the usual 
DGLAP-CS kernels. In HERWIRI, these parameters in principle
can be taken arbitrarily close to zero, as the IR-improved kernels are integrable~\cite{irdglap1,irdglap2}. We note that in the current version of 
HERWIRI, the formula for $\alpha_s(Q)$ is unchanged from that in 
HERWIG6.5 so that there is still a Landau pole therein and this 
would prevent our taking the attendant IR cut-off parameters arbitrarily close to zero; however, we also note that this Landau pole is spurious 
and a more realistic behavior for $\alpha_s(Q)$ as $Q\rightarrow 0$ 
from either the lattice approach~\cite{bou} or from other approaches 
such as those in Refs.~\cite{shirk,max} could be introduced in the 
regime where the usual formula for $\alpha_s(Q)$ fails and this 
would allow us to approach zero with the IR cut-off parameters. With
this understanding,
we now illustrate the difference in IR-cut-off response 
by comparing it for HERWIG6.5 and HERWIRI:
we change the default values of the parameters in HERWIG6.5 by 
factors of 0.7 and 1.44 as shown in the Fig.~\ref{fighw8}. 
We see that the harder cut-off reduces the phase space  
significantly only for the IR-improved kernels and that the softer 
cut-off has also a small effect on the usual kernels' spectra 
whereas as expected
the IR-improved kernels' spectra move significantly toward 
softer values as a convergent integral
would lead one to expect; for, one must note here that the spectra 
all stop at approximately the same value $z_0\cong .00014-.0016$ 
which is above some of the modulated IR-cut-off parameters 
and that the peaks in the spectra are not 
at the
respective IR cut-off values which are defaulted~\cite{herwig} at $z= 
0.000114,\; 0.000121$ for quarks and gluons, respectively, 
as the HERWIG environment has other 
built-in cut-offs that prevent such things as the 
$\alpha_s$ argument's becoming 
too small. What the curves in Fig.~\ref{fighw8} show then are 
the relative ``relative'' probabilities for normalized spectra above $z_0$.
Specifically, the areas under the six curves in the figure are all equal.
The curves then show the difference in the shapes of the parton energy spectra
for the given values of the IR cut-off parameters when the interplay with
the HERWIG environment's other built-in cut-offs to prevent
unphysical phenomena from occurring in the simulation is taken into account.
We can make an estimate
of the attendant relative relative probabilities as follows. We compare
the probability $P(z_0<z<z_1)$ in a normalized spectrum for the spectra
with the IR cut-offs in the usual case, where for soft gluon quanta 
we have $dP\sim dz'/z'$ in
usual DGLAP-CS theory for the IR singular part, with the analogous
probability in the IR-improved case, where for soft gluon quanta we have 
$dP\sim dz'/{z'}^{1-\gamma}$ with $\gamma=\gamma_A,\; A=q,\; G$. 
Considering the $q$ case for definiteness, we have, for $z_0\cong 0.00015$,
an IR cut-off\cite{herwig}  $z_{cq}=0.80\text{GeV}/7\text{TeV}\cong 0.000114$
as noted above,
so that we need to compare the two relative probabilities 
\begin{equation}
P(z_0<z_q<z_1)=
\begin{cases}
\frac{\ln(1-z_0)-\ln(1-z_1)}{\ln(1-z_{cq})-\ln(z_{cG})}, &\text{unimproved}\\
\frac{(1-z_0)^\gamma-(1-z_1)^\gamma}{(1-z_{cq})^\gamma-{z_{cG}}^\gamma}, &\text{IR-improved}
\end{cases}\label{enh1}
\end{equation}
where we also introduced notation for 
the HERWIG environment gluon IR cut-off~\cite{herwig} $z_{cG}=0.85\text{GeV}/7\text{TeV}\cong 0.000121$ with its default value.
This shows that the two relative probabilities are in the
ratio $r\cong (1-{z_{cG}}^\gamma)/(-\gamma\ln(z_{cG}))$ so that the probability
is enhanced in the IR-improved spectra. Putting in the numbers for
we get $r\cong 0.16$ for the suppression of the unimproved spectrum
relative to the IR-improved one. This is of course an overestimate 
of the effect because
we only analyzed the IR singular terms in the soft regime but we see that
the suppression effect is consistent with the plots in Fig.~\ref{fighw8}.
If we want to be more complete, we also need to analyze 
the $\bar{q}$ and $G$ singular
cases and to take the suppression of the soft gluon spectrum in 
the IR-improved spectrum into account. Concerning the $\bar{q}$ case, 
it is the same as the $q$ case -- we get an enhancement of the soft region 
in the IR-improved spectrum relative
to the unimproved one. For the gluon case, when the branching 
is $G\rightarrow G(z)+G(1-z)$, for the part of the gluon branching 
kernel that is singular in $z$
there is an enhancement in the relative probability that the gluon 
with fraction $1-z$ will be in the soft region; the converse also holds. 
Finally, all of the soft gluon 
singularities are suppressed in the IR-improved cases relative
to their unimproved cases, so that this will move particles 
away from the soft regime in the IR-improved cases relative to 
the unimproved cases. 
What is crucial to the Fig.~\ref{fighw8} results is the interplay 
with the HERWIG6.5 environmental cut-offs such as that one on the 
argument of $\alpha_s=\alpha_s(Qz(1-z))$ where $Q$ is the HERWIG6.5 
angle-ordered evolution variable.
This cut-off on the variable ${\cal Q}=z(1-z)Q$ means that the
regime where either $z\rightarrow 0$ or $1-z\rightarrow 0$ is suppressed
in both sets of spectra. But, as the unimproved spectra have their 
largest enhancements in this regime relative to the IR-improved spectra, 
the HERWIG6.5 environment kills a large part of this enhancement 
and allows the other enhancements such as that in (\ref{enh1}) 
to prevail, as we see in the figure.
The behavior illustrated in Fig.~\ref{fighw8}
is expected to lead to a better description of the soft
radiation data at LHC and this expectation should 
be checked with experiment in the not-too-distant future.
\par\indent

\begin{figure*}[t]
%\begin{center}
%%%%\epsfig{file=fig01.ps}
%\epsfig{file=en_sgam-200.eps,width=140mm,height=130mm}
%\end{center}
%\vspace{ -8mm}
%\baselineskip=7mm
\centering
\setlength{\unitlength}{0.1mm}
%%%%%%%%%%%%%%%%%%%%%%%%%%%%%%%%%%
%%%\begin{picture}(1600, 1540)
\begin{picture}(1600, 930)
%%\put( 450, 1530){\makebox(0,0)[cb]{\bf (a)} }
%%\put(1230, 1530){\makebox(0,0)[cb]{\bf (b)} }
%%\put(   0, 870){\makebox(0,0)[lb]{\epsfig{file=ef.ps,angle=270,
%%                                        width=80mm}}}
%%\put( 800, 870){\makebox(0,0)[lb]{\epsfig{file=hwptc.ps,angle=270,
%%                                        width=80mm}}}
%%%\put( 250, 660){\makebox(0,0)[cb]{\bf (a)} }
%%%\put(800, 660){\makebox(0,0)[cb]{\bf (b)} }
%%%\put(   0, 100){\makebox(0,0)[lb]{\epsfig{file=vgcut.eps,angle=90,
%%%                                        width=55mm}}}
%%%\put( 550, 100){\makebox(0,0)[lb]{\epsfig{file=vgcutir.eps,angle=90,
%%%                                        width=55mm}}}
\put( 350, 850){\makebox(0,0)[cb]{\bf (a)} }
\put(1300, 850){\makebox(0,0)[cb]{\bf (b)} }
\put(   -160, 0){\makebox(0,0)[lb]{\includegraphics[width=100mm]{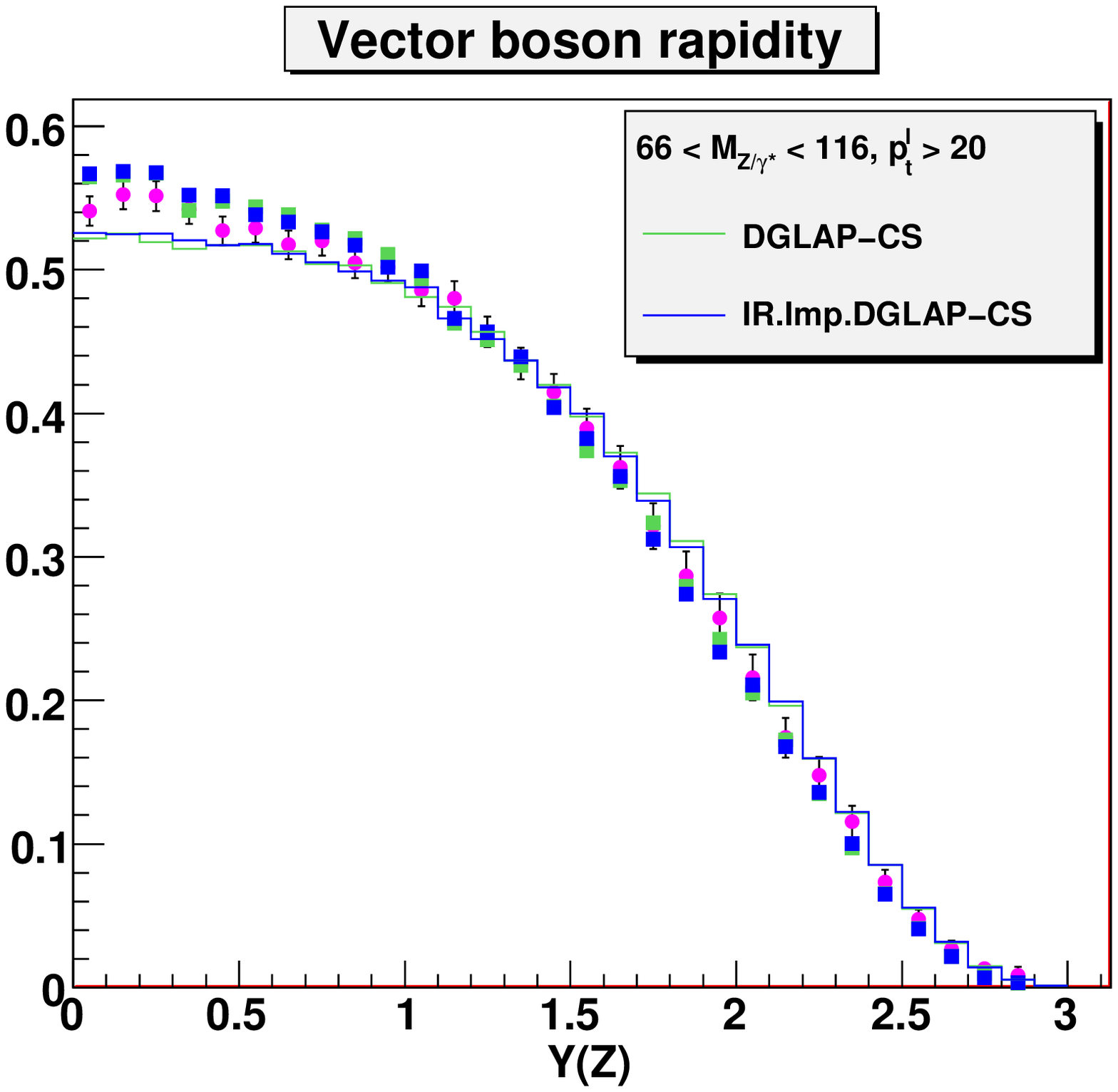}}}
%{\epsfig{file=rapcdf1.ps,angle=90,
%                                        width=80mm}}}
%\put( 820, 0){\makebox(0,0)[lb]{\includegraphics[width=100mm,angle=90]{zptd0spwrd1.eps}}}
\put( 820, 0){\makebox(0,0)[lb]{\includegraphics[width=100mm]{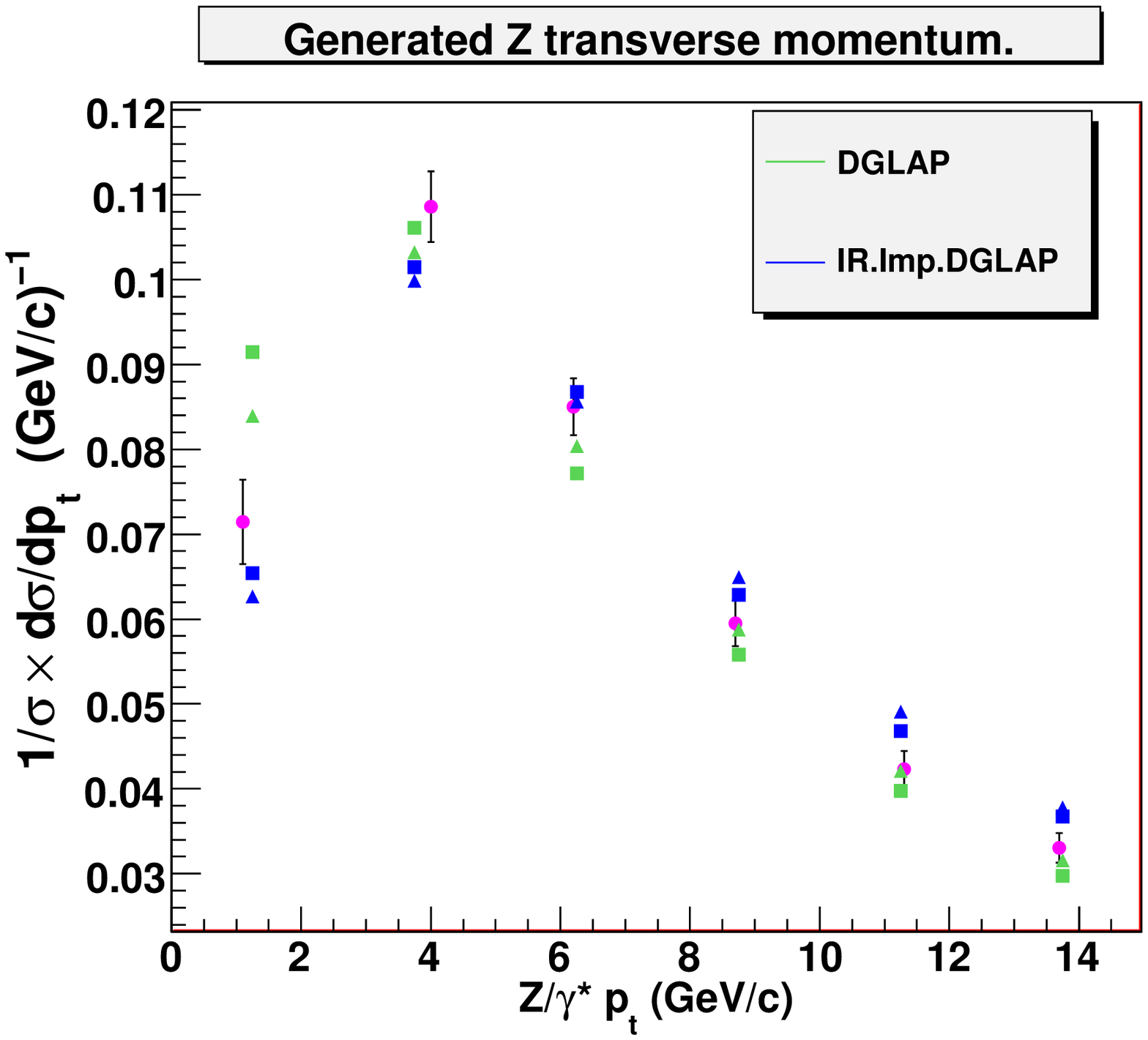}}}
%{\epsfig{file=ptplot090609.ps,angle=90,
%                                        width=80mm}}}
\end{picture}
%\vspace{ -1.5mm}
\caption{Comparison with FNAL data: (a), CDF rapidity data on
($Z/\gamma^*$) production to $e^+e^-$ pairs, the circular dots are the data, the green(blue) lines are HERWIG6.510(HERWIRI1.031); 
(b), D0 $p_T$ spectrum data on ($Z/\gamma^*$) production to $e^+e^-$ pairs,
the circular dots are the data, the blue triangles are HERWIRI1.031, the green triangles are HERWIG6.510. In both (a) and (b) the blue squares are MC@NLO/HERWIRI1.031, and the green squares are MC@NLO/HERWIG6.510, where we use the
notation (see the text) MC@NLO/X to denote the realization by MC@NLO of the exact ${\cal O}{(\alpha_s)}$ correction for event generator X. Note that these are untuned theoretical results.
%\centerline{\Color{PineGreen}COMPARISON WITH DATA IMMINENT.} 
}
\label{fighw9}
\end{figure*}

We finish this initial comparison discussion by turning to the data from 
FNAL on the $Z$ rapidity and $p_T$ spectra as reported in 
Refs.~\cite{galea,d0pt}. We show these results, for 1.96TeV cms 
energy, in Fig.~\ref{fighw9}. We see that HERWIRI1.0(31)
and HERWIG6.5 both give a reasonable 
overall representation of the CDF rapidity data but that
HERWIRI1.031 is somewhat closer to the data for small values of $Y$. The two $\chi^2$/d.o.f are 1.77 and 1.54
for HERWIG6.5 and HERWIRI1.0(31) respectively. The data 
errors in Fig.~\ref{fighw9}(a)
do not include luminosity and PDF errors~\cite{galea}, so that
they can only be used conditionally at this point. We note as well that 
including the NLO
contributions to the hard process via MC@NLO/HERWIG6.510
and MC@NLO/HERWIRI1.031\cite{mcnlo}\footnote{We thank S. Frixione for
helpful discussions with this implementation.} improves the agreement for both
HERWIG6.5 and for HERWIRI1.031, where the $\chi^2$/d.o.f are changed
to 1.40 and 1.42 respectively.
That they are both consistent with
one another and within 10\% of the data in the low $Y$ region is
fully consistent with what we expect given
our comments about the errors and the generic accuracy of
an NLO correction in QCD. A more precise discussion at the NNLO
level with DGLAP-CS IR-improvement and
a more complete discussion  of the errors will
appear~\cite{elswh}. These rapidity
comparisons are then
important cross-checks on our work.
%The two $\chi^2$/p.d.f. are xxx and xxx for HERWIRI1.0 and HERWIG6.5 respectively. 
%The two
%respective $\chi^2$/d.o.f. are xxx and yyy for  HERWIG6.5 and HERWIRI1.0
%for all the data in Fig.~\ref{fighw9}(b).
%The two attendant
%$\chi^2$/p.d.f. are xxx and xxx for HERWIRI1.0 and HERWIG6.5 respectively.
\par\indent

%\begin{figure}
%\begin{center}
%\epsfig{file=vguct.ps,height=60mm}
%\end{center}
%\caption{\baselineskip=7mm     The z-distribution(ISR parton energy fraction) shower comparison in HERWIG6.5 -- preliminary results.}
%\label{hw1}
%\end{figure}
%{\Color{Brown} SIMILAR RESULTS FOR PYTHIA and MC@NLO IN PROGRESS.}
%\item {\Color{PineGreen}COMPARISON WITH DATA IMMINENT.}
%}
%\end{itemize}
\noindent
We also see that HERWIRI1.031 gives a better fit to
the D0 $p_T$ data
compared to HERWIG6.5 for low $p_T$, 
(for $p_T<12.5$GeV, the $\chi^2$/d.o.f. are
$\sim$ 2.5 and 3.3 respectively if we add the statistical and systematic
errors), showing that the IR-improvement makes a better representation
of QCD in the soft 
regime for a given fixed order in perturbation theory.
We have also added
the results of MC@NLO~\cite{mcnlo}
%\footnote{We thank S. Frixione for helpful discussion on this implementation.} 
for the two programs and we see
that the ${\cal O}(\alpha_s)$ correction improves the $\chi^2$/d.o.f for
the HERWIRI1.031 in both the soft and hard regimes and it improves
the HERWIG6.510 $\chi^2$/d.o.f for $p_T$ near $3.75$ GeV
where the distribution peaks. For $p_T<7.5$GeV the $\chi^2$/d.o.f for
the MC@NLO/HERWIRI1.031 is 1.5 whereas that for MC@NLO/HERWIG6.510 is 
worse.
These results are of course still subject to tuning as we indicated above. 
\par\indent

\section{\label{concl} Conclusions}
We have introduced a new approach to QCD parton shower MC's
based on the new IR-improved DGLAP-CS kernels in 
Refs.~\cite{irdglap1,irdglap2} and 
we have realized the new approach with the MC 
HERWIRI1.0(31) in the HERWIG6.5 environment. 
HERWIRI1.0(31) then
sets the stage for
the further implementation
of the attendant~\cite{qced} 
new approach to precision QED$\otimes$QCD predictions for LHC physics
by the introduction of the
respective resummed residuals needed to improve systematically
the precision tag to the 1\% regime for such processes as single heavy gauge boson production, for example. Here, we already note that our new
IR-improved MC, HERWIRI1.0(31), available at http://thep03.baylor.edu, is expected to allow for a better $\chi^2$ per degree of 
freedom in data analysis of high energy
hadron-hadron scattering for soft radiative effects. By comparison with
the D0 FNAL data of single $Z$ production, we have given 
evidence that this is indeed the case. 
As one would expect, the integration of HERWIRI into MC@NLO is seamless,
as one may replace HERWIG with HERWIRI directly. In both cases,
 MC@NLO/HERWIG and MC@NLO/HERWIRI, we see an improvement 
in the comparison with both the CDF rapidity data and the D0 $p_T$ data
on single $Z$ production. 
We await further tests of the new
approach, both at FNAL and at LHC.
\section*{Acknowledgments}
One of us (B.F.L.W) acknowledges helpful discussions with Prof. Bryan Webber
and Prof. M. Seymour and with Prof. S. Frixione. B.F.L. Ward also thanks Prof. L. Alvarez-Gaume and Prof. W. Hollik for the support and kind hospitality of the CERN TH Division and of the Werner-Heisenberg Institut, MPI, Munich, respectively, while this work was in progress. S. Yost acknowledges the hospitality and support of Princeton University and a grant from The Citadel Foundation.

Work partly supported by US DOE grant DE-FG02-09ER41600 and 
by NATO Grant PST.CLG.980342.

%%%STARTHERE

\bigskip
\begin{thebibliography}{99}
\bibitem{jadach1} See for example S. Jadach {\it et al.}, in {\it Geneva 1995, Physics at LEP2, vol. 2}, pp. 229-298; preprint hep-ph/9602393, for a discussion of technical and physical precision.
\bibitem{qedeffects} S. Haywood, P.R. Hobson, W. Hollik and Z. Kunszt,
in {\it Proc. 1999 CERN Workshop on Standard Model Physics ( and more )
at the LHC, CERN-2000-004}, eds. G. Altarelli and M.L. Mangano,( CERN,
Geneva, 2000 ) p. 122
%\bibitem{spies} 
; H. Spiesberger, {\it Phys. Rev. D} {\bf 52} (1995) 4936
%\bibitem{james1} 
; W.J. Stirling,''Electroweak Effects in Parton Distribution 
Functions'', talk presented at ESF Exploratory Workshop,
{\it Electroweak Radiative Corrections to Hadronic Observables at TeV
Energies }, Durham, Sept., 2003
%\bibitem{roth} 
; M. Roth and S. Weinzierl,{\it Phys. Lett. B} {\bf 590} (2004) 190;
J. Blumlein and H. Kawamura, {\it Nucl. Phys. B} {\bf 708} (2005) 467; {\it Acta Phys. Pol. B} {\bf 33} (2002) 3719; 
%\bibitem{james2}
W. J. Stirling et al., 
in {\it Proc. ICHEP04}, eds. H. Chen {\it et al.}
(World Sci. Publ., Singapore, 2005) p. 527; A.D. Martin {\it et al.}, {\it Eur. Phys. J. C} {\bf 39} (2005) 155, and references therein. 
\bibitem{radcor-ew} A. Kulesza {\it et al.}, in {\it PoS RADCOR2007}:001, 2007;
A. Denner {\it et al.}, {\it ibid.}: 002, 2007; A. Denner {\it et al.},{\it  Nucl. Phys. B} {\bf 662} (2003) 299;
G. Balossini et al., arXiv:0805.1129,
and references therein.
%\bibitem{lhclum1} M. Dittmar, F. Pauss and D. Zurcher,{\it Phys. Rev.} D{\bf 56}, 7284 (1997); M. Rijssenbeek, in {\it Proc. HCP2002}, ed. M. Erdmann,( 
%Karlsruhe, 2002 )p. 424; M. Dittmar, {\it ibid.},p.431, and references therein.
\bibitem{ditt-lp09} S. Dittmaier, in {\it Proc. LP09}, 2009, in press.
\bibitem{qced} C. Glosser, S. Jadach, B.F.L. Ward and S.A. Yost,{\it Mod. Phys. Lett. A}{\bf 19}(2004) 2113; B.F.L. Ward, C. Glosser, S. Jadach and S.A. Yost, in {\it Proc. DPF 2004}, {\it Int. J. Mod. Phys. A} {\bf 20} (2005) 3735; in {\it Proc. ICHEP04, vol. 1}, eds. H. Chen {\it et al.},(World. Sci. Publ. Co., Singapore, 2005) p. 588; B.F.L. Ward and S. Yost, preprint BU-HEPP-05-05, in {\it Proc. HERA-LHC Workshop}, CERN-2005-014; in {\it  Moscow 2006, ICHEP, vol. 1}, p. 505; {\it Acta Phys. Polon. B} {\bf 38} (2007) 2395; arXiv:0802.0724, in {\it PoS RADCOR2007}: 038, 2007; B.F.L. Ward {\it et al.}, arXiv:0810.0723, in {\it Proc. ICHEP08}; arXiv:0808.3133, in {\it Proc. 2008 HERA-LHC Workshop},DESY-PROC-2009-02, eds. H. Jung and A. De Roeck, (DESY, Hamburg, 2009)pp. 180-186, and references therein.
\bibitem{jad-ward}  S. Jadach and B.F.L. Ward, {\it Comput. Phys. 
Commun.} {\bf 56}(1990) 351; {\it Phys. Lett. B} {\bf 274} (1992) 470; 
S. Jadach et al., {\it Comput. Phys. Commun.} {\bf 102}
(1997) 229; S. Jadach, W. Placzek and B.F.L Ward, {\it Phys. Lett. B} {\bf 390} (1997) 298; S. Jadach, M. Skrzypek and B.F.L. Ward,{\it Phys. Rev. D} {\bf 55} (1997) 1206; S. Jadach, W. Placzek and B.F.L. Ward, {\it Phys. Rev. D} {\bf 56} (1997) 6939; 
S. Jadach, B.F.L. Ward and Z. Was,{\it Phys. Rev. D} {\bf 63} (2001) 113009;
{\it Comp. Phys. Commun.} {\bf 130} (2000) 260;
S. Jadach et al., {\it ibid.}{\bf 140} (2001) 432, 475.
\bibitem{yfs} D.~R.~Yennie, S.~C.~Frautschi, and H.~Suura, {\it Ann. Phys.} {\bf 13} (1961) 379;\newline
see also K.~T.~Mahanthappa, {\it Phys.~Rev.}~{\bf 126} (1962) 329, for a related analysis.
\bibitem{herwig} G. Corcella {\it et al.}, hep-ph/0210213; {\it J. High Energy Phys.} {\bf 0101} (2001) 010; G. Marchesini et al., {\it Comput. Phys. Commun.}{\bf 67} (1992) 465.
\bibitem{irdglap3-plb} S. Joseph {\it et al.}, {\it Phys. Lett. B}{\bf 685} (2010) 283; 
arXiv:0910.0491.
\bibitem{cattrent} G. Sterman,{\it Nucl. Phys. B} {\bf 281} (1987) 310; S. Catani and L. Trentadue,
{\it Nucl. Phys. B} {\bf 327} (1989) 323; {\it ibid.} {\bf 353} (1991) 183.
\bibitem{scet} See for example C.W. Bauer, A.V. Manohar and M.B. Wise, {\it Phys. Rev. Lett.} {\bf 91} (2003) 122001; C.W. Bauer {\it et al.}
{\it Phys. Rev. D} {\bf 70} (2004) 034014.
\bibitem{dglap}
G. Altarelli and G. Parisi, {\it Nucl. Phys. B} {\bf 126} (1977) 
298; Yu. L. Dokshitzer, {\it Sov. Phys. JETP} {\bf 46} (1977) 641;
L.~N. Lipatov, {\it Yad. Fiz.} {\bf 20} (1974) 181; V. Gribov and L. Lipatov,
{\it Sov. J. Nucl. Phys.} {\bf 15} (1972) 675, 938; see also J.C. Collins and J. Qiu,
{\it Phys. Rev. D}{\bf 39} (1989) 1398 for an alternative discussion
of DGLAP-CS theory.
\bibitem{cs} C.G. Callan, Jr., {\it Phys. Rev. D}{\bf 2} (1970) 1541; K. Symanzik, 
{\it Commun. Math. Phys.} {\bf 18} (1970) 227, 
and in {\em Springer Tracts in Modern Physics},
{\bf 57}, ed. G. Hoehler (Springer, Berlin, 1971) p. 222; see also
S. Weinberg, {\it Phys. Rev. D}{\bf 8} (1973) 3497. 
\bibitem{irdglap1} B.F.L. Ward, {\it Adv. High Energy Phys.} {\bf 2008} (2008) 682312
; DOI:10.1155/2008/682312.
\bibitem{irdglap2} B.F.L. Ward, {\it Ann. Phys.} {\bf 323} (2008) 2147.
%\bibitem{herwig} G. Corcella {\it et al.}, hep-ph/0210213; {\it J. High Energy Phys.} {\bf 0101} (2001) 010; G. Marchesini et al., {\it Comput. Phys. Commun.}{\bf 67} (1992) 465.
\bibitem{scott1} N.E. Adam {\it et al.}, {\it J. High Energy Phys.} {\bf 0805} (2008) 062.
\bibitem{scott2} N.E. Adam {\it et al.}, {\it J. High Energy Phys.} {\bf 0809} (2008) 133.
%\bibitem{yfs}D.~R.~Yennie, S.~C.~Frautschi, and H.~Suura, {\it Ann. Phys.} {\bf 13} (1961) 379; see also K.~T.~Mahanthappa, {\it Phys.~Rev.}~{\bf 126} (1962) 329.
\bibitem{geor1} C. Lee and G. Sterman, {\it Phys. Rev. D} {\bf 75} (2007) 014022.
\bibitem{madg} S.M. Abyat {\it et al.}, {\it Phys. Rev. D} {\bf 74} (2006) 074004.
\bibitem{yellowbook} F. Berends et al., ''Z Line Shape'', in {\it Z Physics at LEP 1, v. 1}, CERN-89-08, eds. G. Altarelli, R. Kleiss, and C. Verzegnassi,(CERN, Geneva, 1989) p. 89, and references therein.
\bibitem{qcdfactorzn} R.K. Ellis {\it et al.}, {\it Phys. Lett. B}{\bf 78} (1978) 281; {\it Nucl. Phys. B} {\bf 152} (1979) 285; D. Amati, R. Petronzio and G. Veneziano, {\it ibid.}{\bf 146} (1978) 29; S.B. Libby and G. Sterman, {\it Phys. Rev. D} {\bf 18} (1978) 3252; A.H. Mueller, {\it ibid.} {\bf 18} (1978) 3705.
\bibitem{ermlv} B.I. Ermolaev, M. Greco and S.I. Troyan, {\it PoS} {\bf DIFF2006} (2006) 036, and references therein.
\bibitem{guido} G. Altarelli, R.D. Ball and S. Forte, 
{\it PoS} {\bf RADCOR2007} (2008) 028.
\bibitem{irdglap4-ichp} B.F.L Ward {\it et al.}, arXiv:0810.0723, 0808.3133.
%\bibitem{dglap}
%G. Altarelli and G. Parisi, Nucl. Phys. {\bf B126} (1977) 
%298; Yu. L. Dokshitzer, Sov. Phys. JETP {\bf 46} (1977) 641;
%L.~N. Lipatov, Yad. Fiz. {\bf 20} (1974) 181; V. Gribov and L. Lipatov,
%Sov. J. Nucl. Phys. {\bf 15} (1972) 675, 938; see also J.C. Collins and J. Qiu,
%Phys. Rev. D{\bf 39} (1989) 1398 for an alternative discussion
%of the lowest order DGLAP-CS theory.
%\bibitem{cs}C.G. Callan, Jr., Phys. Rev. D{\bf 2} (1970) 1541; K. Symanzik, 
%Commun. Math. Phys. {\bf 18} (1970) 227, 
%and in {\em Springer Tracts in Modern Physics},
%{\bf 57}, ed. G. Hoehler (Springer, Berlin, 1971) p. 222; see also
%S. Weinberg, Phys. Rev.D{\bf 8} (1973) 3497; 
%and references therein. 
\bibitem{high-ord-krnls} E.G. Floratos, D.A. Ross, C. T. Sachrajda, {\it Nucl. Phys. B} {\bf 129}(1977) 66;{\it ibid.}{\bf 139}(1978) 545;
{\it ibid.}{\bf 152} (1979) 493,1979; A. Gonzalez-Arroyo, C. Lopez and F.J. Yndurain, {\it Nucl. Phys. B}{\bf 153} (1979) 161; A. Gonzalez-Arroyo and C. Lopez,
{\it Nucl. Phys. B} {\bf 166} (1980) 429; G. Curci, W. Furmanski and R. Petronzio, {\it Nucl. Phys. B} {\bf 175} (1980) 27;  W. Furmanski and R. Petronzio, {\it Phys. Lett. B} {\bf 97} (1980) 437;
E.G. Floratos, C. Kounnas and R. Lacaze, {\it Nucl. Phys. B} {\bf 192} (1981) 417;
R. Hamberg and W. Van Neerven, {\it Nucl. Phys. B} {\bf 379} (1992) 143;
S. Moch, J.A.M. Vermaseren and A. Vogt, {\it Nucl. Phys. B} {\bf 688} (2004) 101; {\it ibid.} {\bf 691} (2004) 129.
\bibitem{hera-dat} See for example A. Cooper-Sarkar {\it et al.} in {\it Proc. 2006 - 2008 HERA-LHC Workshop}, DESY-PROC-2009-02, eds. H. Jung and A. De Roeck,(DESY, Hamburg, 2009) p. 74; T. Carli {\it et al.}, in {\it Proc. HERA-LHC Wkshp}, 2005. 
\bibitem{sac-no-go} C. Di'Lieto, S. Gendron, I.G. Halliday, and C.T. Sachradja, {\it Nucl. Phys. B} {\bf 183} (1981) 223; R. Doria, J. Frenkel and J.C. Taylor, {\it ibid.}{\bf 168} (1980) 93, and references therein.
\bibitem{cat1} S. Catani, M. Ciafaloni and G. Marchesini, {\it Nucl. Phys.} {\bf B264}(1986) 588; S. Catani, {\it Z. Phys.} {\bf C37} (1988) 357.
%\bibitem{qmass} See for example, Particle Data Group (W.-M. Yao {\em et al.}), {\it J. Phys.} {\bf G33} (2006) 1, and references therein.
\bibitem{qmass-bw} B.F.L. Ward, {\it Phys. Rev. D}{\bf 78} (2008) 056001.
\bibitem{baurall}  U. Baur, S. Keller and W.K. Sakumoto, 
{\it Phys. Rev. D} {\bf 57} (1998) 199; U. Baur, S. Keller and D. Wackeroth, {\it ibid.}{\bf 59} (1998) 013002;  U. Baur {\it et al.}, {\it ibid.}{\bf 65} (2002) 033007;
%\bibitem{ditt} 
S. Dittmaier and M. Kramer, {\it Phys. Rev. D} {\bf 65} (2002) 073007; and
%, and references therein
%\bibitem{zyk} 
Z. A. Zykunov,{\it Eur. Phys. J. C}{\bf 3} (2001) 9.
%\bibitem{van1} 
\bibitem{exactqcd}R. Hamberg, W. L. van Neerven and T Matsuura, 
{\it Nucl. Phys. B}{\bf 359} (1991) 343;
%\bibitem{van2} 
W.L. van Neerven and E.B. Zijlstra, {\it Nucl. Phys. B} {\bf 382} (1992)
11; {\it ibid.} {\bf 680} (2004) 513; and 
%\bibitem{anas} 
C. Anastasiou et al., {\it Phys. Rev. D} {\bf 69} (2004) 094008.
%\bibitem{baloss} G. Balossini et al., {\it Nucl. Phys. Proc. Suppl.} {\bf 162} (2006) 59; in {\it Proc. ICHEP06}, eds. A. Sissakian et al. (World Sci. Publ. Co., Singapore, 2008) p. 767. 
%\bibitem{jad-skrz} S. Jadach and M. Skrzypek, in these {\it Proceedings}, 2008;
%{\it Comput. Phys. Commun.} 175 (2006) 511; P. Stevens et al., {\it Acta Phys. Polon.} {\bf B38} (2007) 2379, and references therein.
%\bibitem{cattrent} G. Sterman,{\it Nucl. Phys.}B {\bf 281}, 310 (1987); S. Catani and L. Trentadue,
%{\it Nucl. Phys.}B {\bf 327}, 323 (1989); {\it ibid.} {\bf 353}, 183 (1991).
%\bibitem{scet} See for example C. W. Bauer, A.V. Manohar and M.B. Wise, {\it Phys. Rev. Lett.} {\bf 91} (2003) 122001;
%{\it Phys. Rev.} {\bf D70} (2004) 034014
%%; C. Lee and G. Sterman, {\it ibid.} {\bf D75} (2007) 014022
%, and references therein. 
%\bibitem{herwig} C. Corcella et al., hep-ph/0210213.
\bibitem{pythia} T. Sjostrand et al., hep-ph/0308153.
\bibitem{mcnlo} S. Frixione and B.Webber, {\it J. High Energy Phys.} {\bf 0206} (2002) 029; S. Frixione, P. Nason and B. Webber, {\it ibid.} {\bf 0308} (2003) 007.
\bibitem{skrzjad} S. Jadach and M. Skrzypek, 
{\it Comput. Phys. Commun.} {\bf 175} (2006) 511; P. Stevens et al., {\it Acta Phys. Polon. B} {\bf 38} (2007) 2379, and references therein.
\bibitem{bw-ms-priv-a} We thank  M. Seymour and B. Webber for discussion on this point. 
\bibitem{bw-ms-priv} B. Webber and M. Seymour, private communication.
\bibitem{inprog} M. Hejna et al., to appear.
%\bibitem{scott1} N.E. Adam et al.,{\it J. High Energy Phys.} {\bf 0805} (2008) 062.
%\bibitem{cteq} W.-K. Tung et al., {\it J. High Energy Phys.} {\bf 0702} (2007) 053 and references therein.
%\bibitem{mrst} A.D. Martin et al., {\it Phys. Lett.} {\bf B652} (2007) 292, and references therein.
%\bibitem{fewz} K. Melnikov and F. Petriello, {\it Phys. Rev. Lett.} {\bf 96} (2006) 231803; {\it Phys. Rev.} {\bf D74} (2006) 114017.
%\bibitem{resbos} Q.-H. Cao and C.-P. Yuan, {\it Phys. Rev. Lett.} {\bf 93} (2004) 042001, and references therein.
%\bibitem{horace} C.M. Carloni Calame et al., {\it Phys. Rev.} {\bf D69} (2004) 114017; {\it J. High Energy Phys.} {\bf 0505} (2005) 019;  {\bf 0612} (2006) 016;{\bf 0710} (2007) 109.
%\bibitem{photos} E. Barberio, B. van Eijk and Z. Was, {\it Comput. Phys. Commun.} {\bf 66} (1991) 115;  E. Barberio and Z. Was, {\it ibid.} {\bf 79} (1994) 291; P. Golonka and Z. Was, {\it Eur. Phys. J.} {\bf C45} (2006) 97.
%\bibitem{scott2} N.E. Adam et al., arXiv:0808.0758[hep-ph].
%\bibitem{yfs}D.~R.~Yennie, S.~C.~Frautschi, and H.~Suura, {\it Ann. Phys.} {\bf 13} (1961) 379;\newline
%see also K.~T.~Mahanthappa, {\it Phys.~Rev.}~{\bf 126} (1962) 329, for a related analysis.
%\bibitem{jad-ward}  S. Jadach and B.F.L. Ward, {\it Comput. Phys. 
%Commun.} {\bf 56}(1990) 351; {\it Phys.Lett.} {\bf B274} (1992) 470; 
%S. Jadach et al., {\it Comput. Phys. Commun.} {\bf 102}
%(1997) 229; S. Jadach, W. Placzek and B.F.L Ward, {\it Phys. Lett.} {\bf B390} (1997) 298; S. Jadach, M. Skrzypek and B.F.L. Ward,{\it Phys. Rev. D} {\bf 55} (1997) 1206; S. Jadach, W. Placzek and B.F.L. Ward, {\it Phys. Rev. D} {\bf 56} (1997) 6939; 
%S. Jadach, B.F.L. Ward and Z. Was,{\it Phys. Rev. D} {\bf 63} (2001) 113009;
%{\it Comp. Phys. Commun.} {\bf 130} (2000) 260;
%S. Jadach et al., {\it ibid.}{\bf 140} (2001) 432, 475.
\bibitem{herpp} M. Bahr {\it et al.}, arXiv:0812.0529 and references therein.
\bibitem{cteq} F. Olness, private communication; P.M. Nadolsky {\it et al.}, arXiv:0802.0007.
\bibitem{mrst} R. Thorne, private communication; A.D. Martin {\it et al.}, arXiv:0901.0002 and references therein.
\bibitem{bw-ann-rev} B.R. Webber, {\it Ann. Rev. Nucl. Part. Sci.} {\bf 36} (1986) 253-286.
\bibitem{sjosback} T. Sjostrand, {\it Phys. Lett. B}{\bf 157}(1985) 321.
\bibitem{elswh} S. Joseph {\it et al.}, to appear.
\bibitem{similar} 
Note that similar results for PYTHIA and MC@NLO are in progress.
\bibitem{bou} P. Boucaud {\it et al.}, {\it Nucl. Phys. B Proc. Suppl.} {\bf 106} (2002) 266; J. Skullerud, A. Kizilersu and A.G. Williams, {\it ibid.} {\bf 106} (2002) 841, and references therein.
\bibitem{shirk} M. Baldicchi {\it et al.}, {\it Phys. Rev. Lett.} {\bf 99} (2007) 242001; D.V. Shirkov and I.L. Solovtsov, {\it ibid.} {\bf 79} (1997) 1209; R. Alkofer and L. von Smekal, {\it Phys.\ Rept.} {\bf 353} (2001) 281, and references therein. 
\bibitem{max} P.M. Brooks and C.J. Maxwell, {\it Phys.\ Rev. D} {\bf 74} (2006) 065012, and references therein.
\bibitem{galea} C. Galea, in {\it Proc. DIS 2008}, London, 2008,\newline 
\verb$http://dx.doi.org/10.3360/dis.2008.55$.
\bibitem{d0pt} V.M. Abasov {\it et al.}, {\it Phys. Rev. Lett.} {\bf 100}, 102002 (2008).
%\bibitem{qcdref} B.F.L. Ward and S. Jadach, {\it Acta Phys. Polon.} 
%{\bf B33} (2002)
%1543; in {\it Proc. ICHEP2002},
%ed. S. Bentvelsen et al.,( North Holland, Amsterdam, 2003 ) p. 275
%; B.F.L. Ward and S. Jadach, {\it Mod. Phys. Lett.}{\bf A14} (1999) 491
%;D.B. DeLaney, S. Jadach, C. Shio, G. Siopsis, B.F.L. Ward, {\it Phys. Lett.}{\bf B342}(1995) 239
%; D. DeLaney et al., {\it Mod. Phys. Lett.} {\bf A12} (1997) 2425;
%D. DeLaney et al., {\it Phys. Rev.} {\bf D52} (1995) 108; {\it Phys. Lett.} {\bf B342} (1995) 239; {\it Phys. Rev.} {\bf D66} (2002) 019903(E).
%\bibitem{geor1} C. Lee and G. Sterman, {\it ibid.} {\bf D75} (2007) 014022.
%\bibitem{madg} M. Abyat {\it et al.}, Phys. Rev. {\bf D74} (2006) 074004.
%\bibitem{annphys08} B.F.L. Ward, {\it Ann. Phys.} {\bf 323} (2008) 2147.
%\bibitem{high-ord-krnls} E.G. Floratos, D.A. Ross, C. T. Sachrajda, {\it Nucl.Phys.} {\bf B129}(1977) 66;{\it ibid.}{\bf B139}(1978) 545;
%{\it ibid.}{\bf B152} (1979) 493,1979; A. Gonzalez-Arroyo, C. Lopez and F.J. Yndurain, {\it Nucl. Phys.}{\bf B153} (1979) 161; A. Gonzalez-Arroyo and C. Lopez,
%{\it Nucl. Phys.} {\bf B166} (1980) 429; G. Curci, W. Furmanski and R. Petronzio, {\it Nucl. Phys.} {\bf B175} (1980) 27;  W. Furmanski and R. Petronzio, {\it Phys. Lett.} {\bf B97} (1980) 437;
%E.G. Floratos, C. Kounnas and R. Lacaze, {\it Nucl. Phys.} {\bf B192} (1981) 417;
%R. Hamberg and W. Van Neerven, {\it Nucl. Phys.} {\bf B379} (1992) 143;
%S. Moch, J.A.M. Vermaseren and A. Vogt, {\it Nucl. Phys.} {\bf B688} (2004) 101; {\it ibid.} {\bf B691} (2004) 129, and references therein.
%\bibitem{carli} See for example T. Carli et al., in {\it Proc. HERA-LHC Wkshp}, 2005.
%\bibitem{baurall}  U. Baur, S. Keller and W.K. Sakumoto, 
%{\it Phys. Rev.} D {\bf 57} (1998) 199; U. Baur, S. Keller and D. Wackeroth, {\it ibid.}{\bf 59} (1998) 013002;  U. Baur {\it et al.}, {\it ibid.}{\bf 65} (2002) 033007, and references therein.
%\bibitem{ditt} S. Dittmaier and M. Kramer, {\it Phys. Rev.} D{\bf 65} (2002) 073007, and references therein
%\bibitem{zyk} Z. A. Zykunov,{\it Eur. Phys. J.}C{\bf 3} (2001) 9, and references therein.
%\bibitem{van1} R. Hamberg, W. L. van Neerven and T Matsuura, 
%{\it Nucl. Phys.}B{\bf 359} (1991) 343.
%\bibitem{van2} W.L. van Neerven and E.B. Zijlstra, {\it Nucl. Phys.} B{\bf 382} (1992)
%11; {\it ibid.} B{\bf 680} (2004) 513; and, references therein.
%\bibitem{anas} C. Anastasiou et al., {\it Phys. Rev.} D{\bf 69} (2004) 094008.
%\bibitem{baloss} G. Balossini et al., {\it Nucl. Phys. Proc. Suppl.} {\bf 162} (2006) 59; in {\it Proc. ICHEP06}, eds. A. Sissakian et al. (World Sci. Publ. Co., Singapore, 2008) p. 767. 
%\bibitem{jad-skrz} S. Jadach and M. Skrzypek, in these {\it Proceedings}, 2008;
%{\it Comput. Phys. Commun.} 175 (2006) 511; P. Stevens et al., {\it Acta Phys. Polon.} {\bf B38} (2007) 2379, and references therein.
%\bibitem{lee-naun} T.D. Lee and M. Nauenberg, 
%{\it Phys. Rev.} 133 (1964) B1549.
%\bibitem{elsewh} B.F.L. Ward et al., to appear.
%\bibitem{hypgeo} M. Yu. Kalmykov, B.F.L. Ward and S.A. Yost, {\it J. High Energy Phys.} {\bf 0702} (2007) 040; {\it ibid.} {\bf 0710} (2007) 048; {\it ibid.} {\bf 0711} (2007) 009; S.A. Yost,  M. Yu. Kalmykov and B.F.L. Ward, in {\it Proc. ICHEP08}, arXiv:0808.2605; M. Yu. Kalmykov and
%B.A. Kniehl, arXiv:0807.0567, and references therein.
\end{thebibliography}
\end{document}